\documentclass[12pt,preprint]{aastex}

\usepackage{epsfig,subfigure}

\newcommand{\etal}{et~al.\ }

\newcommand{\feka}{\hbox{Fe\,K$\alpha$}}

\newcommand{\simgt}{\lower 2pt \hbox{$\, \buildrel {\scriptstyle >}\over {\scriptstyle\sim}\,$}}
\newcommand{\simlt}{\lower 2pt \hbox{$\, \buildrel {\scriptstyle <}\over {\scriptstyle\sim}\,$}}

\newcommand{\chandra}{{\emph{Chandra}}}

\newcommand{\ie}{i.e.,\,}
\newcommand{\be}{\begin{equation}}
\newcommand{\ee}{\end{equation}}
\newcommand{\bea}{\begin{eqnarray}}
\newcommand{\eea}{\end{eqnarray}}

\newcommand{\dtglens}{1.17^{+0.41}_{-0.31}  \times 10^{-22}\rm mag\,cm^2\,atom^{-1}}

\newcommand{\mdtg}{0.54^{+0.19}_{-0.14}  \times 10^{-22}\rm mag\,cm^2\,atom^{-1}}
\newcommand{\dtgGRB}{0.39^{+0.19}_{-0.14}  \times 10^{-22}\rm mag\,cm^2\,atom^{-1}}

\shorttitle{\emph{CHANDRA} OBSERVATIONS of GRAVITATIONAL LENSES}
\shortauthors{CHEN ET AL.}

\tighten

\begin{document}

\def\sarc{$^{\prime\prime}\!\!.$}
\def\arcsec{$^{\prime\prime}$}
\def\arcmin{$^{\prime}$}
\def\degr{$^{\circ}$}
\def\seco{$^{\rm s}\!\!.$}
\def\ls{\lower 2pt \hbox{$\;\scriptscriptstyle \buildrel<\over\sim\;$}}
\def\gs{\lower 2pt \hbox{$\;\scriptscriptstyle \buildrel>\over\sim\;$}}

\title{Dust, Gas, and Metallicities of Cosmologically Distant Lens Galaxies }

\author{Bin Chen\altaffilmark{1}, Xinyu Dai\altaffilmark{1}, Christopher S. Kochanek\altaffilmark{2}, George Chartas\altaffilmark{3} }

\altaffiltext{1}{Homer L. Dodge Department of Physics and Astronomy, The University of Oklahoma,
Norman, OK, 73019, USA, bchen@ou.edu}

\altaffiltext{2}{Department of Astronomy, The Ohio State University, Columbus, OH 43210, USA}

\altaffiltext{3}{Department of Physics and Astronomy, College of Charleston, SC 29424, USA}

\begin{abstract}
We homogeneously analyzed the \chandra\ X-ray observations of 10 gravitational lenses,  HE~0047$-$1756, QJ~0158$-$4325, SDSS~0246$-$0805, HE~0435$-$1223, SDSS~0924$+$0219, SDSS~1004+4112, HE~1104$-$1805,  PG~1115+080, Q~1355$-$2257, and Q~2237+0305, to measure the differential X-ray absorption between images, the metallicity, and the dust-to-gas ratio of the lens galaxies.
We detected differential absorption in all lenses except SDSS~0924$+$0219 and HE~1104$-1805$.
This doubles the sample of dust-to-gas ratio measurements in cosmologically distant lens galaxies. 
We successfully measured the gas phase metallicity of three lenses, Q~2237+0305, SDSS~1004+4112, and B~1152$+$199 from the X-ray spectra.
Our results suggest a linear correlation between metallicity and dust-to-gas ratio (i.e., a constant metal-to-dust ratio), consistent with what is found for nearby galaxies.
We obtain an average dust-to-gas ratio $E(B-V)/N_H=\dtglens$ in the lens galaxies, with an intrinsic scatter of $\rm0.3\,dex$.
Combining these results with data from GRB afterglows and quasar foreground absorbers, we found a mean dust-to-gas ratio $\mdtg,$ now significantly lower than the average Galactic value,  $1.7\,\times 10^{-22}\,\rm mag\, cm^{2}\, atoms^{-1}.$
This suggests evolution of dust-to-gas ratios with redshift and lower average metallicities for the higher redshift galaxies, consistent with current metal and dust evolution models of interstellar medium.
The slow evolution in the metal-to-dust ratio with redshift implies very rapid dust formation in high redshift ($z>2$) galaxies.
\end{abstract}


\section{INTRODUCTION}

Understanding the interstellar medium (ISM) and the intergalactic medium (IGM) is important for many fields of modern astronomy, especially for star and galaxy formation and evolution studies (e.g., Madau et al.\ 1998; Steidel et al.\ 1999; Cen \& Ostriker 1999; Dav$\rm\acute{e}$ \& Oppenheimer 2007; Conroy et al.\ 2009).
In addition, extinction by dust in the ISM needs be measured and corrected in order to use high redshift standard candles such as SNe Ia to constrain cosmological models (e.g., Riess et al.\ 1998, Perlmutter et al.\ 1999), or to correctly determine the physical properties of sources such as GRBs (Piran 2004).
Studies of ISM properties start by measuring the extinction or absorption it produces.
In the Milky Way and nearby galaxies ($<10\rm\, Mpc$), extinction and extinction laws can be measured by comparing spectra of reddened and un-reddened stars of the same spectral type (the ``pair" method, see Savage \& Mathis 1979; Massa et al.\ 1983; Cardelli et al.\ 1989; Fitzpatrick 1999; Gordon et al.\ 2003). 
Galactic extinction curves can be reasonably well-modeled by a one-parameter ($R_V\equiv A_V/E(B-V)$) model sequence with $\langle R_V\rangle= 3.1$ although there are differences between sight lines (e.g., those through molecular clouds) and galaxies.    
For example, the mean extinction curve of the LMC shows a much weaker $2175\rm\AA$ feature than the Milky Way and a steeper rise in the far-UV band, and that of SMC shows no $2175\rm\AA$ feature (see Draine 2003 for a review).
The average Galactic dust-to-gas ratio ($\equiv E(B-V)/N_H$, where $N_H$ is the neutral hydrogen column density) was found by Bohlin et al.\ (1978) to be $1.7\times 10^{-22}\,\hbox{mag}\, {\rm cm}^{2}\,{\rm atoms}^{-1}$ with an intrinsic scatter of $\sim$30\%.
The dependences of the dust-to-gas ratio on Galactocentric radius and its relation to the metallicity gradient of the Milky Way and nearby galaxies were investigated by Issa et al.\ (1990), who found a linear correlation between the dust-to-gas ratio and metallicity, implying a constant metal-to-dust ratio. 
The physical form of the extinction law thus depends on the metallicity, the ionization state, the mean size, composition and alignment of the dust grains of the ISM, and all these are expected to be function of redshift.

The dust, metal, and gas ratios can also be used to test ISM formation and evolution models. 
Several groups have carried out detailed simulations of the evolution of the ISM including its dust-to-gas ratio, metallicity, and dust-to-metal ratio (e.g., Dwek 1998; Edmunds 2001; Inoue 2003). 
Generally in these models, dust particles are formed both in quiescent stellar outflows and supernova ejecta, they grow by accretion from the gas phase and coagulation with other dust particles, and are destroyed by evaporation and collisions. 
The evolution of the dust-to-gas ratio can be estimated because the surface density of gas is directly related to the star formation rate, which in turn is related to the supernova and stellar mass-loss rates. 
Since dust is formed around stars in the late stages of stellar evolution and star formation regulates the supply of gas, the dust-to-gas ratio should be related to the star formation rate, with lower dust-to-gas ratios for galaxies with higher star formation rates. 
Since the star formation rate has dropped dramatically from $z\sim1$ to the present day, we expect a lower dust-to-gas ratio in high redshift galaxies. 
Indeed, this picture is supported by simulations of the evolution of the ISM (Edmunds 2001; Inoue 2003).

However, it is much more difficult to measure extinction/absorption properties of high redshift galaxies. 
In particular, the ``pair" method does not work for galaxies at cosmological distances because it is impossible to obtain accurate photometry of individual stars. 
For high redshift galaxies, extinction curves have been studied using SNe Ia (Riess et al.\ 1996, Perlmutter  et al.\ 1997), quasars with damped $\rm Ly\alpha$ (DLA) systems in the foreground (Pei et al.\ 1991; Murphy \& Liske 2004; York et al.\ 2006; $\rm\ddot{O}$stman et al.\ 2006; M$\rm\acute{e}$nard et al.\ 2008, 2009), and afterglows of gamma-ray bursts (GRBs; Price et al.\ 2001; Jha et al.\ 2001; Jakobsson et al.\ 2004; Kann et al.\ 2006; Watson et al.\ 2013; Covino et al.\ 2013; Zafar \& Watson 2013; De Cia et al.\ 2013).
These methods generally rely on having an accurate determination of the intrinsic continuum, which is difficult in many cases.
For example, $\rm\ddot{O}$stman et al.\ (2006) suggest an evolution of dust properties with redshift, with lower values of $R_V$ in Ly$\alpha$ systems.
Jakobsson et al.\ (2004) found that the extinction of the optical afterglow of the GRB~030429 is best fit with a SMC-like extinction (see also Schady et al.\ 2007).
Covino et al.\ (2013) compiled a sample of GRBs complete in redshift and found no clear redshift evolution in the dust extinction properties of the afterglows.
They found a dust-to-gas ratio $\sim$$0.20\times 10^{-22}\rm\,mag\,cm^2\,atoms^{-1}$ in the GRB host galaxies, about 10\% the Galactic value, consistent with Stratta et al.\ (2004) and Schady et al.\ (2010).
Zafar \& Watson (2013) computed the metals-to-dust ratios of the ISM in a wide variety of galaxies by combining samples of GRB afterglows, quasar foreground absorbers, and gravitational lenses (Dai \& Kochanek 2009), and found that the metals-to-dust ratios in these systems are surprisingly close to the Galactic value.  
They found no evidence for any dependence of the mean metals-to-dust ratio on galaxy type or age, redshift, or metallicity, and their results favor Supernovae ejecta as the dominant  formation mechanism for the bulk of dust in the ISM. 

Gravitationally lensed quasars provide an independent method to measure extinction and extinction laws in high redshift ($z>0$) galaxies (Nadeau et al.\ 1991).   
Multiply lensed quasars intersect the lens galaxy at different locations, and are sensitive to any differential absorption/extinction between the quasar images.
By measuring the differential absorption/extinction between the quasar images, we can measure the ISM properties of the lens without the prior knowledge of the intrinsic continuum of the quasars. 
This method has been applied in the optical bands to single systems by several groups (Nadeau et al.\ 1991; Jaunsen \& Hjorth 1997; Toft et al.\ 2000; Motta et al.\ 2002; Wucknitz et al.\ 2003; Mu$\rm\tilde{n}$oz et al.\ 2004; Mediavilla et al.\ 2005) and to multiple lenses in surveys (Falco et al.\ 1999; El$\rm\acute{\i}$asdottir et al.\ 2006; Mu$\rm\tilde{n}$oz et al.\ 2011).
Using 23 gravitational lens systems Falco et al.\ (1999) found that the extinction in the lenses is patchy and shows no correlation with impact parameter.  
El$\rm\acute{\i}$asdottir et al.\ (2006) studied the extinction properties of 10 lens systems, and found no strong evidence for evolution in extinction properties with redshift. 
Mu$\rm\tilde{n}$oz et al.\ (2011) constructed UV extinction curves covering the $2175\rm\AA$ feature for three lenses, HE~0512$-$3329, B~1600+434, and H~1413+117.
In X-rays, differential absorption column densities $\Delta N_H$ have been measured for several lenses by Dai et al.\ (2003, 2006), Dai \& Kochanek (2005, 2009).
Dai \& Kochanek (2009) studied the evolution of the dust-to-gas ratio of ISM using a sample of 8 gravitational lenses, finding an average dust-to-gas ratio $E(B-V)/N_H=(1.5\pm0.5)\times 10^{-22}\,\rm mag\, cm^2\, atoms^{-1}$ with an intrinsic scatter of $\sim$40\%.
Both the average value and the intrinsic scatter are found to be consistent with the Galactic values (Bohlin et al.\ 1978). 
Since the X-ray absorption cross section is dominated by metals, and the $N_H$ is inferred assuming a solar metallicity, these results can also be interpreted as implying that the metal-to-dust ratios in cosmologically distant galaxies are consistent with the solar value.

In this paper, we analyze \chandra\ observations of  10 gravitational lenses, HE~0047$-$1756 (Wisotzki et al.\ 2004; Ofek et al.\ 2006), QJ~0158$-$4325 (Maza et al.\ 1995; Morgan et al.\ 1999; Faure et al.\ 2009), SDSS~0246$-$0805 (Inada et al.\ 2005; Eigenbrod et al.\ 2006), HE~0435$-$1223 (Wisotzki et al.\ 2002; Ofek et al.\ 2006), SDSS~0924+0219 (Inada et al.\ 2003a; Ofek et al.\ 2006), SDSS~1004+4112 (Inada et al.\ 2003b), PG~1115+080 (Weymann et al.\ 1980; Kundi$\rm\acute{c}$ et al.\ 1997),  Q~1355$-$2257 (Morgan et al.\ 2003; Eigenbrod et al.\ 2006), HE~1104$-$1805 (Wisotzki \etal 1993; Lidman et al.\ 2000), and Q~2237+0305 (Huchra et al.\ 1985), to measure the differential X-ray absorption between the lensed images, the metallicity, and the dust-to-gas ratio of lens galaxies. 
For the highest S/N spectra we also measure the gas phase metallicities for comparison with ISM evolution models.  
The basic data for the 10 lenses is given in Table~\ref{tab:lens-info}.

\section{OBSERVATIONS AND DATA REDUCTION}

\begin{figure*}
\begin{center}$
\begin{array}{cc}
\includegraphics[width=0.45\textwidth,height=0.32\textheight]{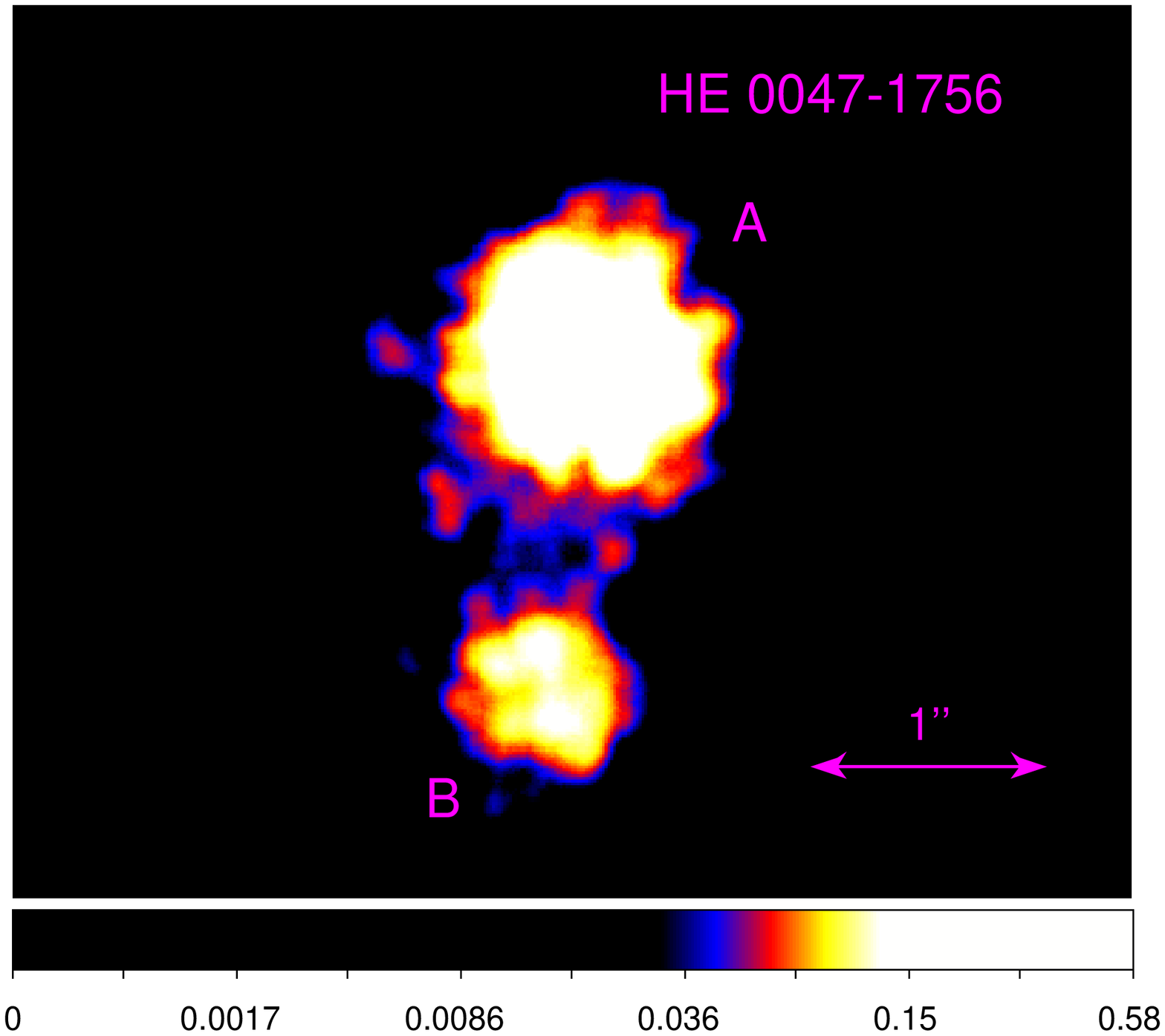}
\hspace{15pt}
\includegraphics[width=0.45\textwidth,height=0.32\textheight]{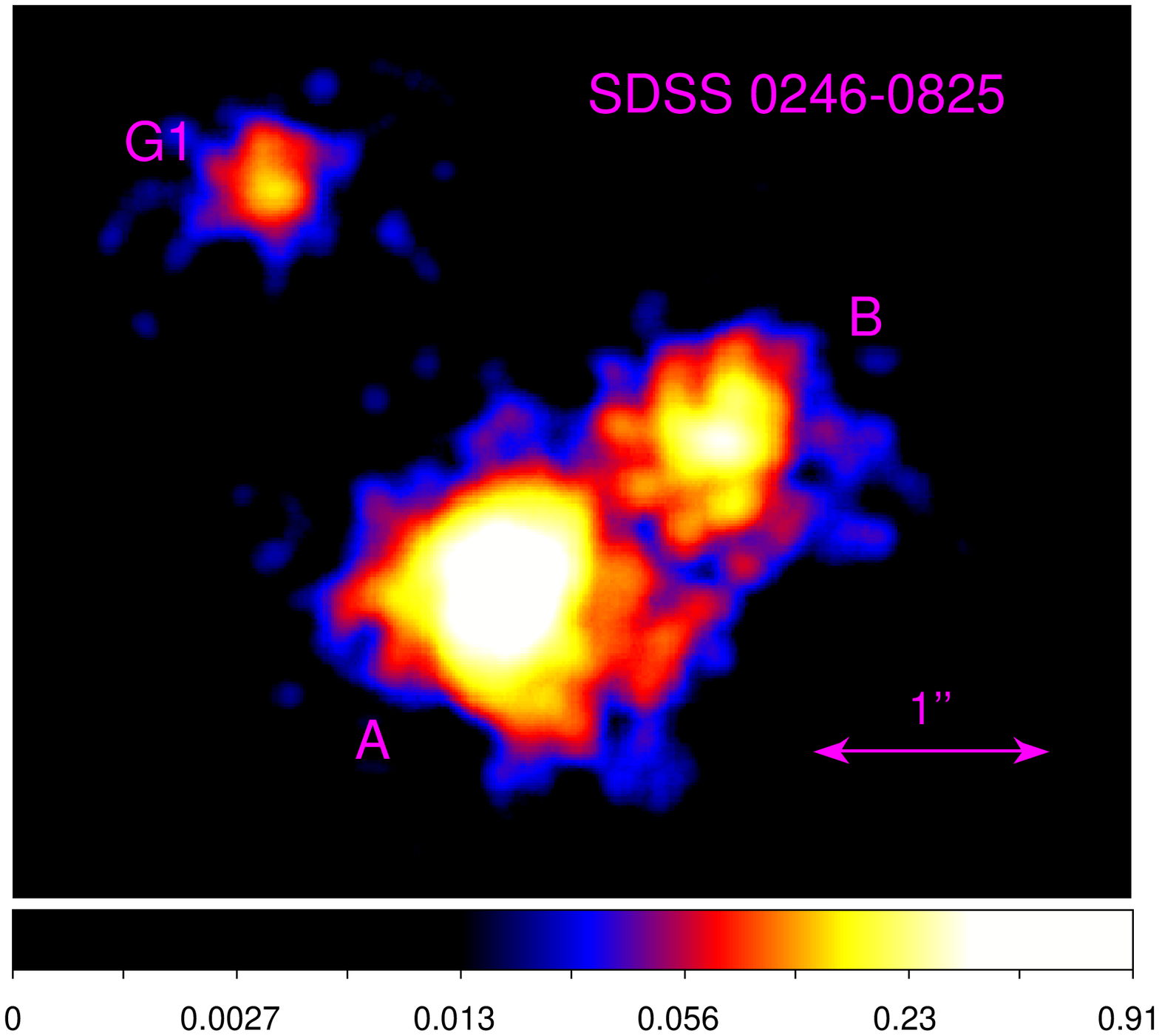}\\
\vspace{15pt}
\includegraphics[width=0.45\textwidth,height=0.32\textheight]{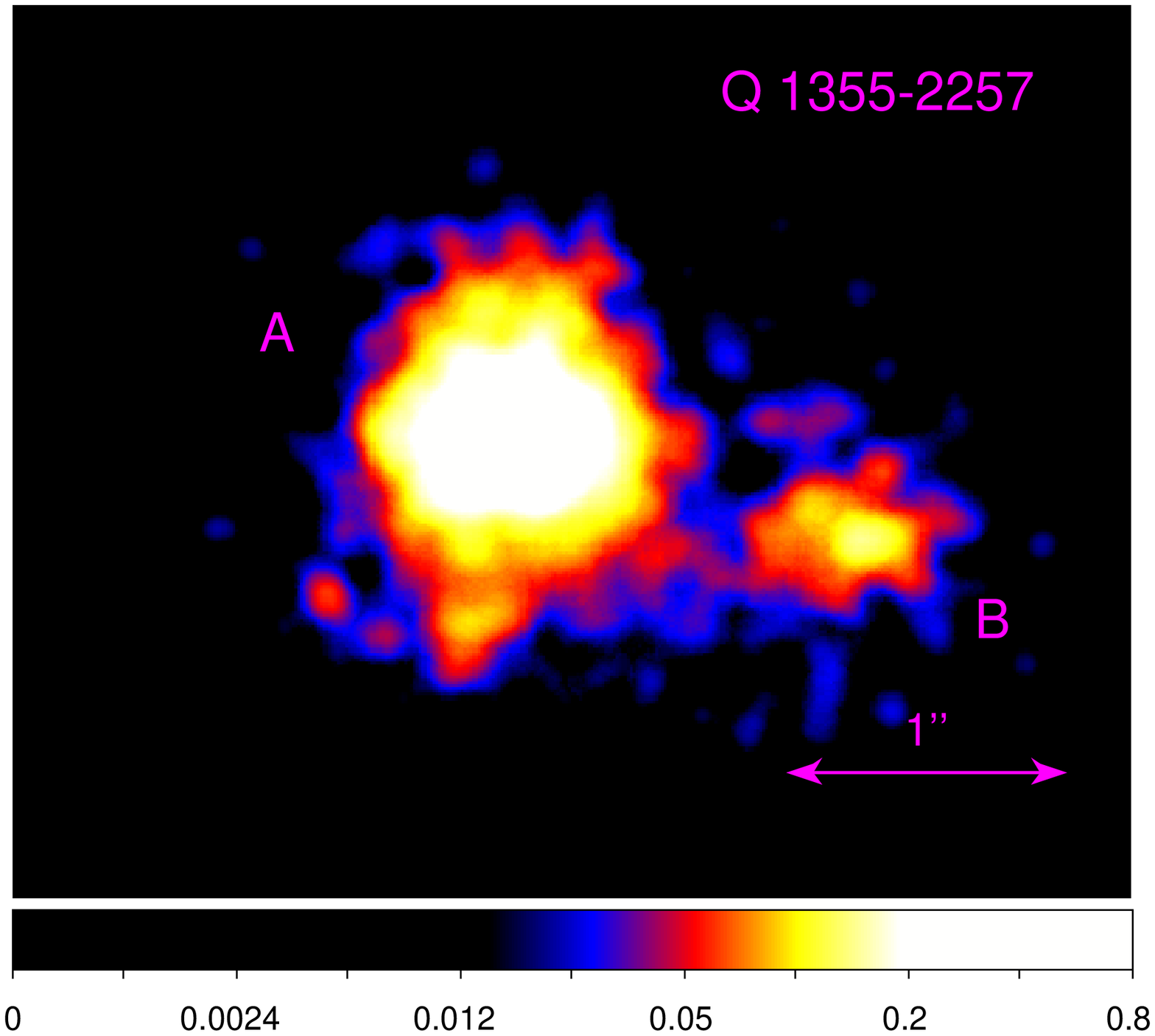}
\hspace{15pt}
\includegraphics[width=0.45\textwidth,height=0.32\textheight]{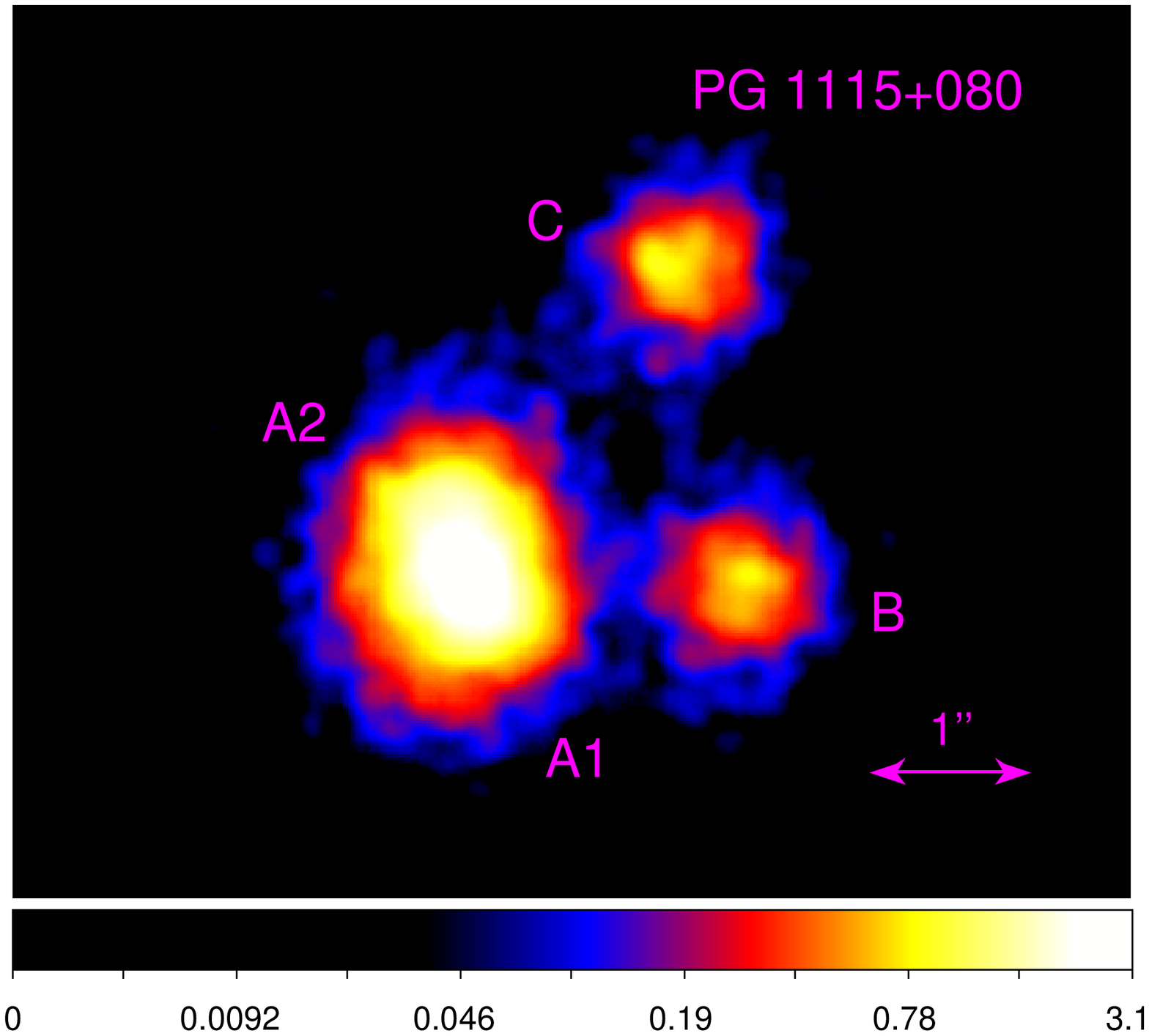}
\end{array}$
\end{center}
\caption{\chandra\ X-ray images of HE~0047$-$1756, SDSS~0246$-$0825, Q~1355$-$2257, and PG~1115$+$080.
	      In the top right panel, the object labeled by $G1$ is a nearby quasar (not the lens galaxy which is between the image A and the image B).
	      The images of HE~0047$-$1756 and PG~1115$+$080 combine two and nine epochs, respectively.
	      \label{fig:stacked_image}}
\end{figure*}

All X-ray observations were made with the Advanced CCD Imaging Spectrometer (Garmier et al.\ 2003) on board \chandra\ (Weisskopf et al.\ 2002).
The majority of the observations are from our \chandra\ Cycle 11 large program to study quasar microlensing.
Three observations are from our \chandra\ Cycle 9 program to study dust-to-gas ratios of lens galaxies, and the remaining are archival observations. 
A log of the new \chandra\ observations is given in Table~\ref{tab:chandra_new}.
Observations of QJ~0158$-$4325, HE~0435$-$1223, SDSS~0924+0219, SDSS~1004+4112, HE~1104$-$1805, and Q~2237+0305 were reported in Chen et al.\  (2012).
The \chandra\ data were reduced using CIAO 4.3 software tools provided by the \chandra\ X-ray Center (CXC).
Since the lensed images are closely separated, we reprocessed all data to remove the standard pixel randomization and instead apply a sub-pixel assignment algorithm to the event files (Li et al.\ 2004). 
We used only events with standard $ASCA$ grades of 0, 2, 3, 4, and 6 in the analysis. 
Figure~\ref{fig:stacked_image} shows the images of  Q~1355$-$2257, SDSS~0246$-$0805, HE~0047$-$1756, and PG~1115+080. 
The images of HE~0047$-$1756 and PG~1115$+$080 are stacked over two and nine epochs, respectively.
The stacked images of the other six lenses were shown in Figure~1 of Chen et  al.\ (2012). 

\section{SPECTRAL ANALYSIS}

For each lens we extracted the spectra of the individual images using the CIAO tool ``psextract" in circular regions of radii about 0\sarc5--0\sarc8 that contain most of the photon counts.
For the cluster lens SDSS~1004+4112, we used apertures of radii 1\sarc5.
Thanks to the superb angular resolution of \chandra, the X-ray background is dominated by contaminating emission from adjacent images.
Therefore to estimate the contamination of image A by image B, we measured the background spectrum in the circular region symmetric to image A with respect to image B. 
For the cluster lens SDSS~1004+4112, we extracted the background spectra from concentric partial rings at the same radius from the cluster center to model the contamination from the cluster X-ray emission.
Images C and D of SDSS~0924+0219 are faint and poorly resolved, consequently we extracted a single spectrum containing images A, C, and D.
After extracting the spectra of the individual images for each epoch, we obtained the combined spectra for each image by stacking the spectra from different epochs and the corresponding \verb+arf+ and \verb+rmf+ files.
For lenses with more than one observation, we focus on the combined spectra in the spectral analysis.

We model the observed X-ray spectrum of the $i^{\rm th}$ image of the lensed quasar as 
\bea\label{model}
N_i(E,t) &=& N_{0}(t-\Delta t_i)\left(\frac{E}{E_0}\right)^{-\Gamma(t-\Delta t_i)}M_i\left[E(1+z_l),t\right]\cr
&&\times \exp\left\{-\sigma(E)N_H^{\rm Gal}-\sigma\left[E(1+z_l)\right]N_H^i-\sigma\left[E(1+z_s)\right]N_H^{\rm Src}(t-\Delta t_i)\right\}.  
\eea 
The model consists of a source power law in energy magnified by $M_i$ and absorbed by gas in the source galaxy, the lens at image position $i,$ and the Milky Way, with hydrogen column densities of $N_H^{\rm Src}$, $N_H^i,$ and $N_H^{\rm Gal},$ respectively. 
In Eq.~(\ref{model}) $\Delta t_i$ is the time delay of image $i,$ $\Gamma$ is the power-law index of the source spectrum, and $\sigma(E)$ is the photoelectric absorption cross section.
Since the photoelectric absorption cross section decreases with energy ($\sigma(E)\propto E^{-3}$ approximately), the absorption at the source will affect the observed spectrum the least compared with absorption at the lens and the Milky Way for the same $N_H.$
We therefore ignored the source absorption in our analysis except for PG~1115+080 (see section~\ref{sec:solar}).  
The absorption in the Milky Way and the lens are still degenerate after dropping $N_H^{\rm Src}$.  
The Milky Way absorption is the same for all images, so the differential absorption between images of the lens depends only on the ISM of the lens galaxies. 
If there is significant source absorption (i.e., much larger than that in the lens galaxy or the Milky Way), the measured lens absorption column density $N_H^i$ can be significantly biased.
However, we expect the error in the measured differential absorption $\Delta N_H$ between images to be still small, provided that the temporal variation of the source absorption column density is not significant during the time scale corresponding to the lensing time delay.     
We also note that for spectra with low signal-to-noise ratio, the power-law index $\Gamma$ can be partially degenerate with the absorption column density.

In Section~\ref{sec:solar} we fit the X-ray spectra assuming solar abundances for the lens galaxies.
In Section~\ref{sec:vary} we fit the spectra with varying metallicities.
We estimate dust extinction and contamination caused by chromatic microlensing in Section~\ref{sec:dust}.   
We fit the spectra using XSPECV12 (Arnaud 1996).

\subsection{Spectral Fits Assuming Solar Metallicity in the Lens Galaxies}\label{sec:solar}

In this Section we used the standard $wabs$ and $zwabs$ models in XSPEC for Galactic and lens absorption, respectively, assuming solar abundances (Anders \& Grevesse 1989). 
The Galactic $N_H$ was fixed to values from Dickey \& Lockman (1990) as given in Table~\ref{tab:lens-info}. 
We assumed the same power-law index $\Gamma$ for all images of the background quasar, but assumed that the absorption $N_{H}^i$ in the lens differed for each image.
We also fixed the absorber redshift to the lens redshift.  
The results for the spectral fits are given in Table~\ref{tab:spectra}.
The spectra of QJ~0158$-$4325, HE~0435$-$1223, SDSS~0924+0219, SDSS~1004+4112, HE~1104$-$1805, and Q~2237+0305 were analyzed in detail in Chen et al.\ (2012) and
we simply include those results in Table~\ref{tab:spectra} for completeness.

\begin{figure*}
\begin{center}$
\begin{array}{cc}
\includegraphics[width=0.5\textwidth,height=0.35\textheight]{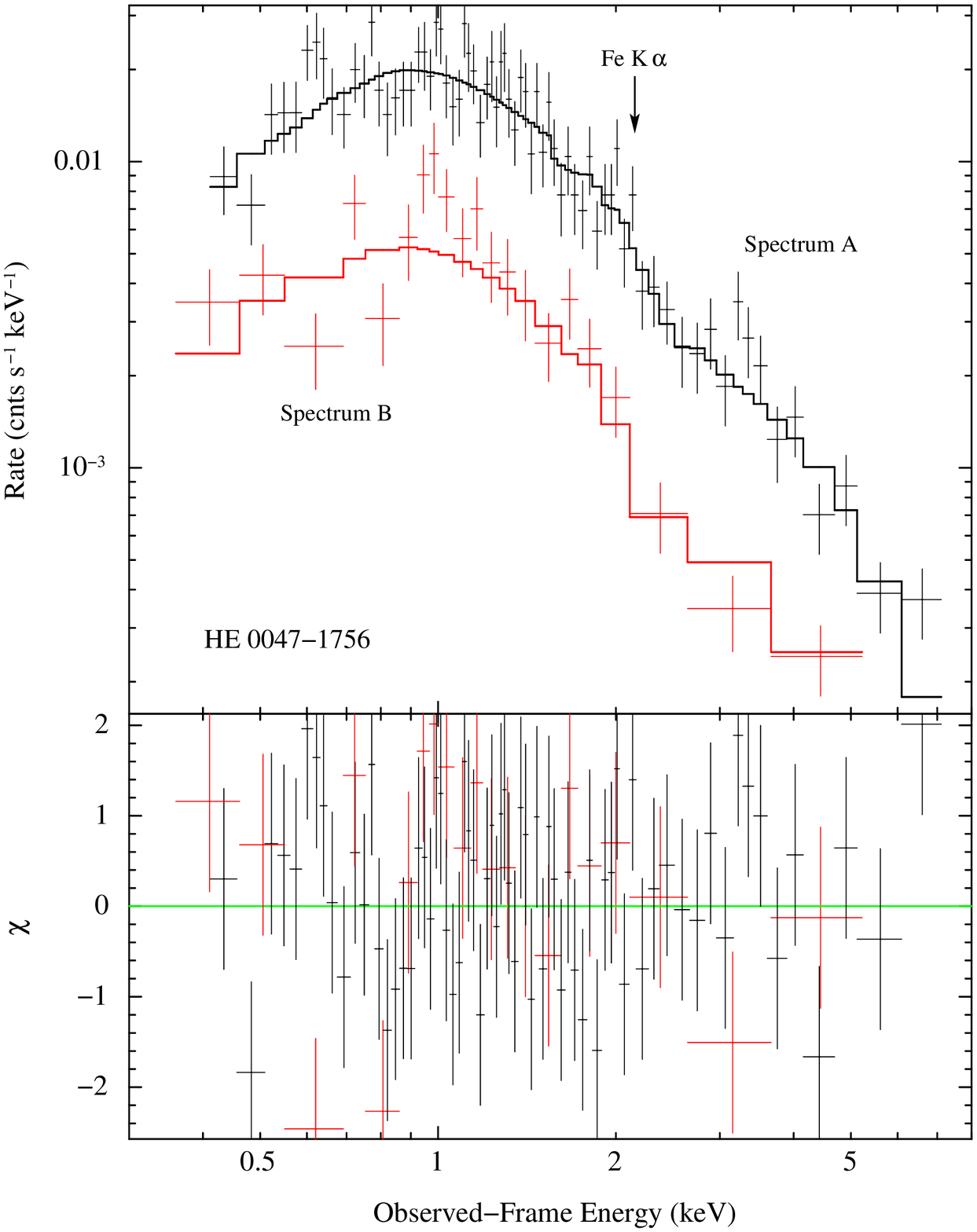}
\hspace{15pt}
\includegraphics[width=0.5\textwidth,height=0.35\textheight]{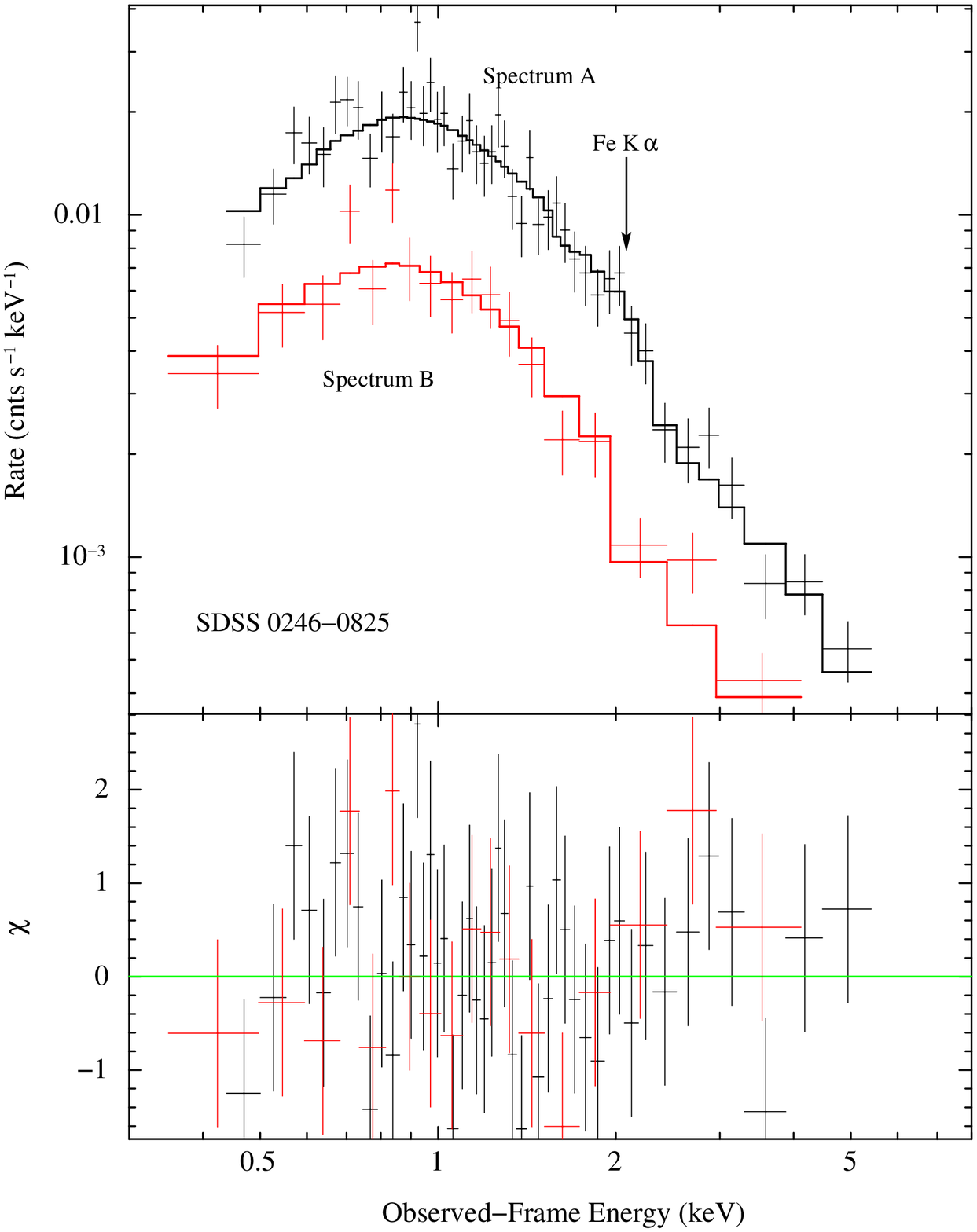}\\
\vspace{15pt}
\includegraphics[width=0.5\textwidth,height=0.35\textheight]{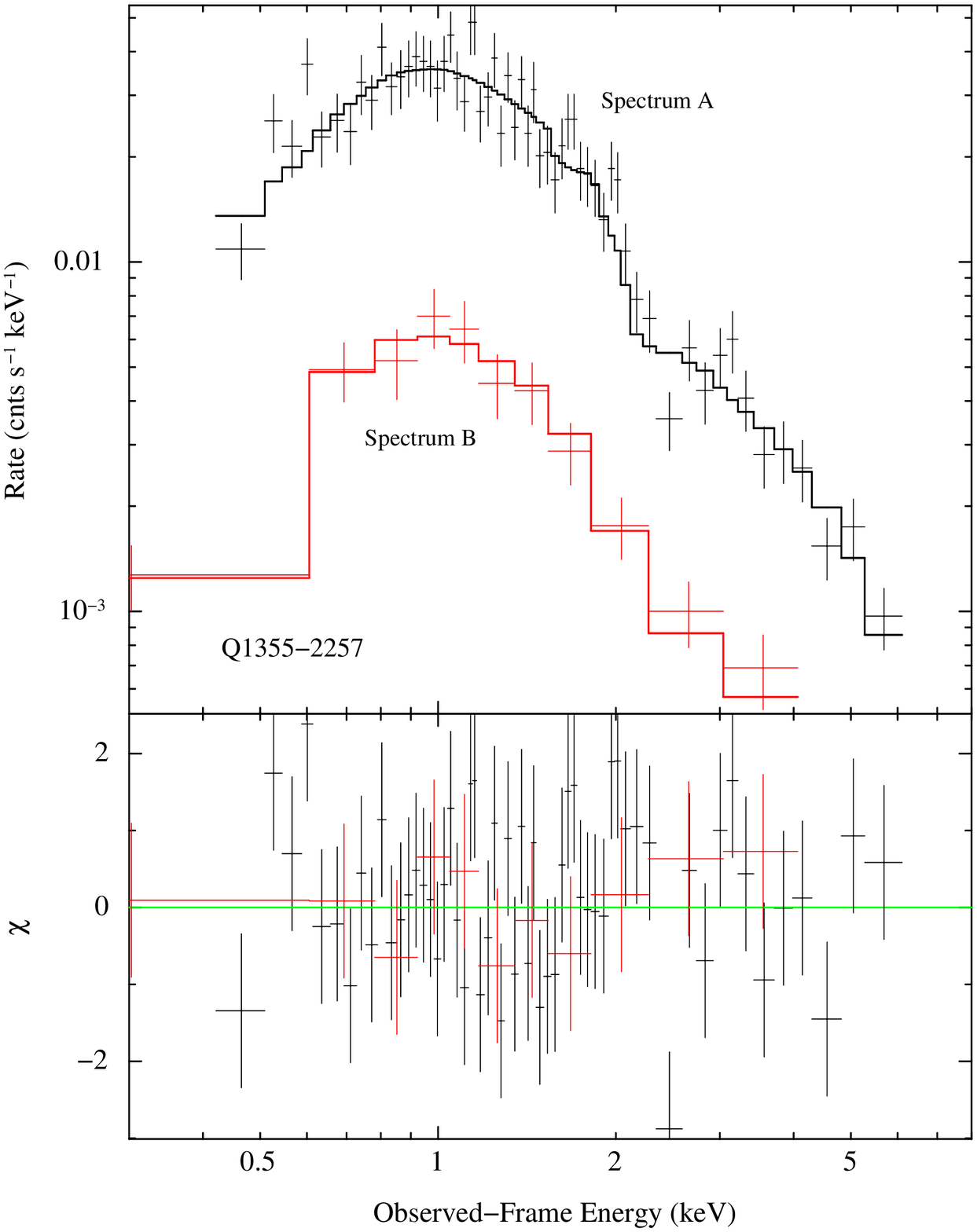}
\hspace{15pt}
\includegraphics[width=0.5\textwidth,height=0.35\textheight]{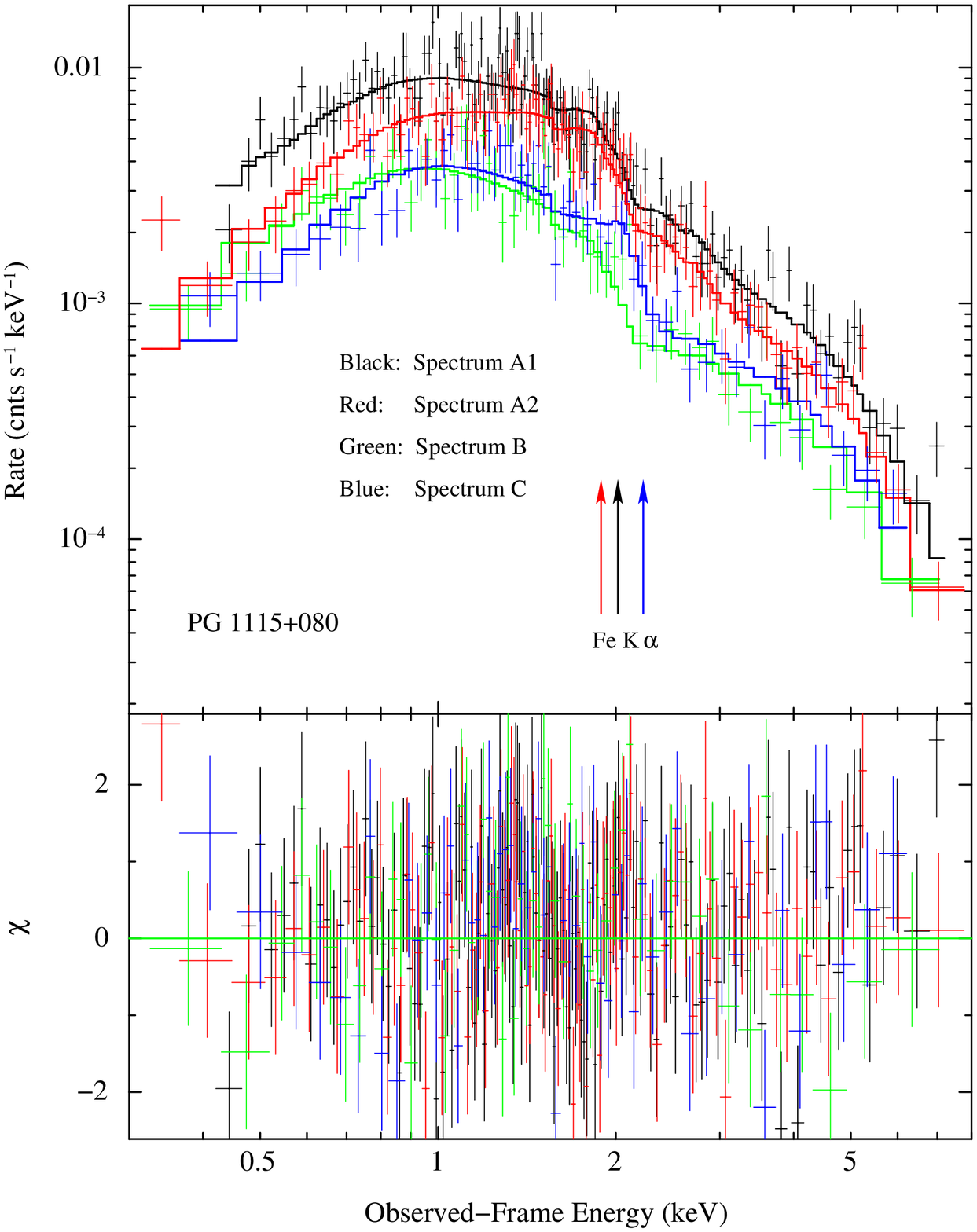}
\end{array}$
\end{center}
\caption{   Spectra and the best fit models of HE~0047$-$1756, SDSS~0246$-$0825, Q~1355$-$2257, and PG~1115$+$080, based on a power-law source spectrum with a Gaussian emission line modified by lens and Galactic absorption. 
		The spectra of HE~0047$-$1756 and PG~1115$+$080 combine two and nine epochs, respectively.
		The \feka\ line was detected for the brighter image A of  HE~0047$-$1756 and SDSS~0246$-$0825, and the images A1, A2, and C of PG~1115$+$080, but not for Q~1355$-$2257.	
	      \label{fig:spectra}}
\end{figure*}

In this paper, we only discuss the fits to Q~1355$-$2257, SDSS~0246$-$0805, HE~0047$-$1756, and PG~1115+080 in detail (Figure~\ref{fig:spectra}). 
Power-law emission modified by foreground absorption gives acceptable fits for Q~1355$-$2257, HE~0047$-$1756, SDSS~0246$-$0825 (reduced $\chi^2_{\nu}=1.09,$ $1.20,$ $1.15$ for $\rm d.o.f.= 65$, 88, and 61, degrees of freedom, respectively), but not for PG~1115$+$080 (a reduced $\chi^2_{\nu}=1.31$ with $\rm d.o.f. = 363$).
The standard disk-corona model for AGN X-ray emission predicts the existence of a reflection component in the observed X-ray spectrum including emission lines such as \feka\ (Reynolds \& Nowak 2003).
This line is detected in almost all lensed quasars with high S/N spectra (Chen et al.\ 2011, 2012; Chartas et al.\ 2012).
The EW of the lines are higher than those unlensed quasars of similar luminosity and the line energy can be blue or redshifted from the rest 6.4 keV due to microlensing effects.
We next fit the spectra of these four lenses by a power law with  Gaussian emission lines (\ie the $zgauss$ model in XSPEC) to test whether we can obtain better fits.
We detected redshifted iron K$\alpha$ line emission (rest frame energy $6.4$ keV) in the brighter image A of the two image lens HE~0047$-$1756 and SDSS~0246$-$0825, and in images A1, A2, and C of the quadruple lens PG~1115$+$080.
We detected no line emission in either image of Q~1355$-$2257.
We summarize these detections in Tables~\ref{tab:iron}.
We used an F-test to test the significance of these lines, and they are all $>95\%$ significant (Table~\ref{tab:iron}).
For example, adding the iron line component improves the fits of PG~1115$+$080 to $\chi^2_\nu=1.05$ with $\rm d.o.f.=354$.
The redshifted lines (observed fame 2.14 keV for the image A of HE~0047$-$1756 and SDSS~0246$-$0825, $1.88$ keV, $2.0$ keV, and $2.1$ keV for the images A1, A2 and C of PG~1115+080) are difficult to visualize because they fall near to the energy range where the instrument response changes rapidly.
Adding a blueshifted emission line component at $3.09$ keV (rest frame energy $7.32$ keV; see Figure~\ref{fig:spectra}) for the image A of Q~1355$-$2257 slightly improves the fit (reduced $\chi^2=1.06$ for $\rm d.o.f.=62$).  
After adjusting the significance of the line detection by the numbers of trials to go from the best fit line center (7.32 keV) to the expected line center ($6.4$ keV), we concluded that the detection is not statistically significant. 
The addition of these lines has little consequence for the $N_H$ measurements because they represent such a small fraction of the total spectrum.

Ignoring source absorption will mainly influence PG~1115+080, a mini broad absorption line (mini-BAL) quasar for which source absorption might be significant (Chartas et al.\ 2003).
If we fit its spectra including source absorption (adding a $zwabs$ component $N_{H}^{Src}$ at the source redshift $z_s$), the fits do not improve and we find $N_H^{\rm Src}<0.08\times10^{22}\,\rm cm^{-2}$ in a joint fit to the spectra of all four images.  
We also fit the spectra of the images A1 and A2 (we chose this pair of images for the dust-to-gas ratio measurement) using either a common or two independent source absorption components and found a upper limit of $N_{H}^{\rm Src}\lesssim0.47\times10^{22}\,\rm cm^{-2}.$
As expected, the change in the differential absorption between images is insignificant (see Table~\ref{tab:spectra}).  
Chartas et al.\ (2003) analyzed the X-ray spectra of PG~1115$+$080 extracted from the \chandra\ and {\emph{XMM-Newton}} observations, and detected two high-energy absorption lines at the rest frame energy 7.4 keV and 9.5 keV, respectively.
They explained these two absorption features as X-ray absorbers (most likely highly ionized Fe ions)  in a relativistic outflow moving with velocities of $\sim 0.10 c$ and $\sim0.34 c,$ respectively.  
They also detected variability in the absorption line energy and  equivalent width by comparing the \chandra\ and \emph{XMM-Newton} observations separated by 19 weeks.
We fit the combined spectra of PG~1115$+$080 (over 9 epochs) using a power-law model with absorption lines at these energies (\ie the standard $gabs$ model of XSPEC) and the fit was not improved significantly.
This is not surprising since the rapid variability of X-ray absorbers in relativistic flows (e.g., Chartas et al.\ 2007) might have erased these features in the combined spectra.
On the other hand, if the iron line emission comes from the reflection of coronal X-ray emission by the accretion disk and is less variable than the absorption line, then the combining spectra should help in detecting the emission lines.
This might explain why we have detected the iron emission line in a much higher signal-to-noise ratio spectrum while not finding the absorption features as found in Chartas et al.\ (2003).  

In general we found that a simple power-law model with emission lines modified by lens and Galactic absorption fits the spectra of all 10 lenses well. 
The best fit power-law index lies between $1.7$ and $2.2,$ typical of quasar X-ray emission (Reeves \& Turner 2000; Dai et al.\ 2004; Saez et al.\ 2008).
We detect differential absorption in 8 out of the 10 gravitational lenses with SDSS~0924+0219 and HE~1104-1805 as the two exceptions.
A marginal differential absorption $\Delta N_H=0.055\pm0.030$ between the image A and the image B of HE~1104$-$1805 was reported in Dai et al.\ (2006).
We detected no absorption for either image of this lens based on combining all the \chandra\ data totaling $\sim$110 ks.

\subsection{Spectral Fits With Varying Metallicity}\label{sec:vary}

For an ISM with elemental abundances similar to the solar value, the photoelectric absorption of soft X-rays is dominated by the inner shell electrons of the most abundant metals (C, N, O, Ne and S) along with the L-shell electrons of Fe (Wilms et al.\ 2000).  
Since solar metallicity is often assumed when measuring the hydrogen content of the ISM using X-ray absorption, the estimate of $N_H$ can be readily transformed into the column density of metals,  and the dust-to-gas ratios computed this way are sometimes called the metal-to-gas ratios (e.g. M$\rm\acute{e}$nard \& Chelouche 2009; Watson et al.\  2013; Zafar \& Watson 2013).
So far, we have assumed solar metallicity $Z_\odot$ for the lenses when fitting the X-ray spectra with absorption models.
However, if the metallicity of the lenses differs from the solar value, the column density (and dust-to-gas ratio) measured through X-ray absorption assuming solar metallicity will then be biased.    
To remove the degeneracy between hydrogen column density and the ISM metallicity, and to test the possible evolution of the metallicity with redshift, we also fit the spectra of each lens with a second lens absorption model, the standard $zvphabs$ model of XSPEC, which allows the metallicity to vary. 
When fitting this model, we fix the abundance of He to be solar, but allow the metallicity to vary relative to solar.  
We ignored source absorption in this subsection.

\begin{figure}
	\epsscale{0.6}
	\plotone{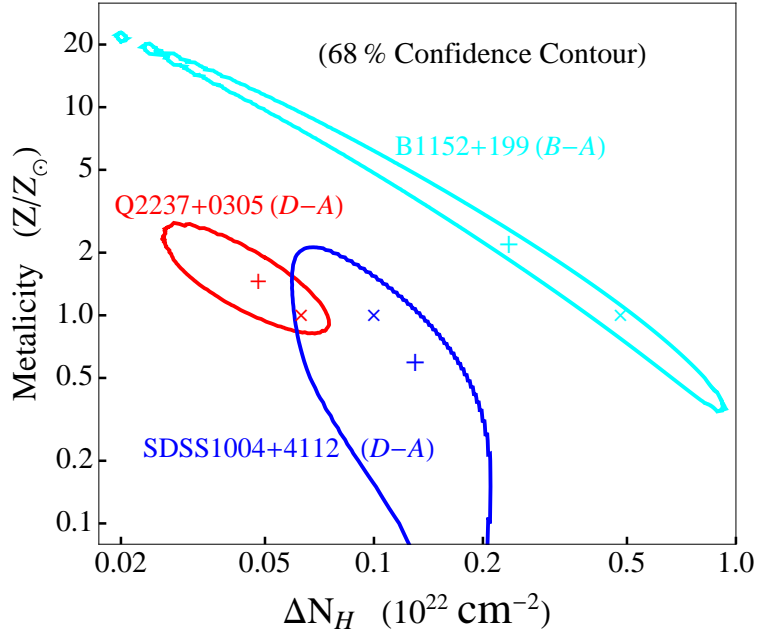}
	\caption{Joint estimates of the differential column density $\Delta N_H$ ($10^{22}\rm cm^{-2}$) and the metallicity $Z/Z_\odot$ in solar units for the lens galaxy Q~2237+0305 (between images D and A), B~1152+199 (between images B and A), and the cluster lens SDSS~1004+4112 (between images D and A).
	The ``$\times$" sign marks the best fit model with fixed $Z/Z_\odot\equiv1$ while the ``$+$" sign marks the best fit when the metallicity is allowed to vary.
	The contours are at the $68\%$ confidence limit for two parameters.
		       	   \label{fig:total_metal}} 
\end{figure}

\begin{figure}
	\epsscale{0.6}
	\plotone{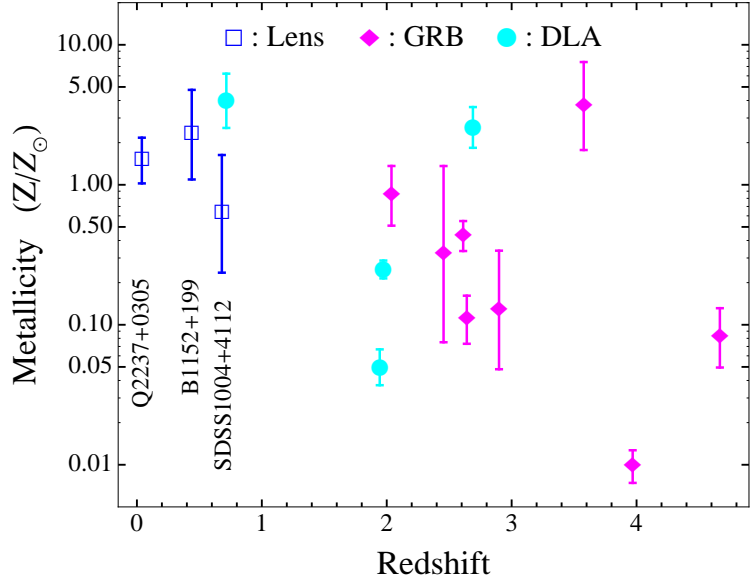}
	\caption{	Estimated lens metallicity $Z/Z_\odot$ in solar units as a function of redshift. 
			The metallicity and redshift data for the GRBs and quasar Ly$\alpha$ absorbers are given in Table~\ref{tab:zafar13}. 
			The metallicities of the three lenses are consistent with the solar value given their uncertainties.
			The metallicities of the GRB sample are lower than the solar value on average (with large scatter). 
		       	   \label{fig:Metal_Z}} 
\end{figure}

The fits allowing varying metallicity converge only for two lenses with long total exposure times, SDSS~1004+4112, and Q~2237+0305.
The metallicities of these two lenses (in units of the solar metallicity $Z_\odot$) are shown in Table~\ref{tab:spectra}.
We also fit this second absorption model to an archival spectrum of B~1152+199 (Dai \& Kochanek 2009) which has a similarly high S/N. 
We are able to independently constrain $N_H$ and the metallicity, with the results given in Table~\ref{tab:spectra}.
For these three lenses, we show the joint confidence contours for the differential absorption $\Delta N_H$ and the metallicity in Figure~\ref{fig:total_metal}. 
The metallicity is still degenerate with the hydrogen column density, especially in B~1152+199.
The best fit metallicities differ from the solar value but are still consistent with it given the uncertainties.
We find no evidence of redshift evolution using this small sample of lenses with direct gas-phase metallicity measurements, as shown in Figure~\ref{fig:Metal_Z}.
SDSS~1004+4112 is a cluster lens, where the metallicities can also be measured using the X-ray emission from the cluster using \chandra.
Ota et al.\ (2006) found $Z=0.21Z_\odot$ (90\% upper limit of $0.62Z_\odot$; no lower limit reported) that is lower but consistent with our absorption estimates.


\subsection{Extinction Estimates and Chromatic Microlensing}\label{sec:dust}

To compute the dust-to-gas ratios of gravitational lenses, we need to measure the differential extinction between the images of the background quasars.  
Specifically, we compute the differential color excess, $\Delta E(B-V)\equiv  (B-V)_i - (B-V)_j$ in the rest frame of the lens (indices $i$ and $j$ label images of the background quasar) with estimates of the differential gas column density.
We use the $\Delta E(B-V)$ values in Falco et al.\ (1999) for HE~1104$-$1805, PG~1115+080, and Q~2237+0305, see Table~\ref{tab:dtg}.
For the other lenses, we had to estimate the differential extinction using available optical band data (Kochanek et al.\ 2007; Wisotzki et al.\ 2004; Inada et al.\ 2005).  
The available data for these lenses do not allow a detailed modeling of the extinction law, we therefore assumed an average $R_V=3.1$ Galactic extinction law when computing $\Delta E(B-V)$ (Cardelli et al.\ 1989).
The differential extinctions of QJ~0158$-$4325, Q~1355$-$2257, SDSS~1004$+$4112, and HE~0435$-$1223 are estimated using the optical/near-IR data from CASTLES (three filters: F160W, F555W, and F814W; Kochanek et al.\ 2007), while those for SDSS~0246$-$0825 and HE~0047$-$1756 are based on the flux ratios from Inada et al.\ (2005) and Wisotzki et al.\ (2004), respectively (both have five filters).   
Since the CASTLES data (Kochanek et al.\ 2007) have magnitude measurements for only three bands, the uncertainties in $\Delta E(B-V)$ are relatively large (Table~\ref{tab:dtg}). 
Observations in the IR/Optical/UV band with multiple filters, in particular, near the rest frame $2175\rm\AA$ feature should allow more accurate measurements of extinction and extinction laws for these high redshift lenses (e.g., Mu$\rm\tilde{n}$oz et al.\ 2011), and significantly improve the measurements of differential extinction and the dust-to-gas ratios. 

The main systematic uncertainty in our results is chromatic microlensing by stars in the foreground lens galaxy (or cluster), which can contaminate the measurements of both the extinction (color excess) and the column density.
If the source size or profile is wavelength dependent and comparable in size to the Einstein radii of the stars in the lens galaxy or smaller (e.g., in a thin accretion disk the characteristic size $R_\lambda\propto \lambda^{4/3}$, Sunyaev \& Shakura 1973), then the observed spectra can be magnified/demagnified by different factors in different energy bands.
Consequently, chromatic microlensing  contaminates the measurement of optical extinction curves, the X-ray spectral index and the absorption column density.
In the optical band, chromatic microlensing has been observed for several systems (e.g., Poindexter et al.\ 2008; Anguita et al.\ 2008; Eigenbrod et al.\ 2008; Bate et al.\ 2008; Blackburne et al.\ 2011a,\ b; Mu$\rm\tilde{n}$oz et al.\ 2011). 
This indeed causes problems in measuring extinction and extinction laws in some lens systems, such as HE~1104$-$1805 (Dai et al.\ 2006; Mu$\rm\tilde{n}$oz et al.\ 2011) where microlensing has led to a reversal of the optical image colors since its discovery. 
While X-ray microlensing has been detected (Chartas et al.\ 2002; Dai et al.\ 2003; Blackburne et al.\ 2006; Pooley et al.\ 2007; Chen et al.\ 2012), chromatic X-ray microlensing was not detected until very recently (Chen et al.\ 2011 in Q~2237+0305 and Chartas et al.\ 2012 in RX~J1131$-1231$).
Since the quasar X-ray emission regions are more compact than the optical emission (Kochanek et al.\ 2004; Chartas et al.\ 2009; Dai et al.\ 2010; Blackburne et al.\ 2011a,\ b, 2013; Morgan et al.\ 2012; Mosquera et al.\ 2013), microlensing should be stronger in the X-ray bands and chromatic between optical and X-ray emission.
Our lenses are chosen to have significant extinction, reducing the importance of microlensing.
Furthermore, we average over many epochs of \chandra\ observations which will reduce the effects of microlensing.
Of the ten lenses reported in this paper, eight were observed more than once (see Tables~\ref{tab:chandra_new}).
Unfortunately, long term multicolor optical light curves are not available for many of these lenses.
A few lenses, such as Q~1355$-$2257 and HE~0047$-$1756, we have only one published optical observation. 

Another potential source of systematic error is the ionization state of the ISM.
We computed the hydrogen column density $N_H$ by fitting the absorbed X-ray spectra assuming photoelectric absorption by a neutral ISM.
We have also assumed solar abundance for all but three lenses.    
If some components of the ISM, particularly the hydrogen, are partially or completely ionized, or the elemental abundances are significantly different from the solar value, then the $N_H$ computed from X-ray spectral fitting can be biased or erroneous.
For example, the ionization state of the X-ray absorbing material is very important for understanding the X-ray absorption in GRB afterglows. 
Watson et al.\ (2013) claimed that the soft X-ray absorption of GRB afterglows is dominated by He in the GRB's host's H~II region rather than by metals (which should be the case assuming neutral ISM with solar abundances) or hydrogen.
Since we are examining random sight lines in the lens galaxy, assuming a largely neutral ISM is reasonable.
This issue could be explained in detail by studying the UV metal absorption (e.g., C~II, O~I, Fe~II, Mg~I, Mg~II) in and between the lensed images.
 

\section{DISCUSSION}

We obtained 6 useful dust-to-gas ratios out of the 10 lenses with new or additional \chandra\ observations. 
We detected no absorption in HE~1104$-$1805 and SDSS~0924$+$0219, and we had to drop QJ~0158$-$4325 and SDSS~0246$-$0825 since the differential extinction $\Delta E(B-V)$ and absorption $\Delta N_H$ are of opposite sign. 
This indicates that microlensing is more important than absorption.
We know QJ~0158$-$4325 is strongly microlensed (Morgan et al.\ 2012) while SDSS~0246$-$0825 is less well studied.
We have a total of 11 dust-to-gas ratios after including the 5 lenses from Dai et al.\ (2006) and Dai \& Kochanek (2009).
This nearly doubles the sample size, and extends the sample to smaller extinctions and column densities.
We also updated the dust-to-gas ratio of B~1152+199 using the new value of differential absorption column density obtained from a spectral fit allowing varying metallicity. 
We summarize the dust-to-gas ratio data of the lenses in Table~\ref{tab:dtg}.
Besides gravitational lenses, we also consider the sample of GRBs and QSO DLAs compiled in Table~1 of Zafar \& Watson (2013), including only those systems with both absorption and extinction measurements (including uncertainty estimates), as listed in Table~\ref{tab:zafar13}.
We converted the $A_V$ values to color excess $E(B-V)$ assuming a Galactic extinction law with $R_V=3.1,$ and the resulting dust-gas-ratios  $N_H/E(B-V)$ are included in Table~\ref{tab:zafar13}. 
For some of the GRBs and DLAs, the metallicity $Z$ of the ISM was estimated using the column density $N_{S}$ or $N_{Zn}$ of Sulfur or Zinc.
We have  28 measurements of dust-to-gas ratios and  15 measurements of metallicities after including GRB and quasar $\rm Ly\alpha$ absorbers.   
We first make a test for correlations between metallicity and dust-to-gas ratios using the three lenses with direct metallicity measurements in Table~\ref{tab:metallicity} and the GRBs and DLAs with metallicity measurements in Table~\ref{tab:zafar13}. 

\subsection{Metallicity and Metal-to-Dust Ratio of Gravitational Lenses, GRB Host Galaxies, and Quasar Foreground Absorbers}

The correlation between the metallicity and dust-to-gas ratio of the ISM was first found more than two decades ago using galaxies in the Local Group such as the Milky Way, the Large and Small Magellanic Clouds, M31, and M33 (Viallefond et al.\ 1982; Issa et al.\ 1990).
The correlation suggests a constant metal-to-dust ratio.
Dwek (1998) presented a detailed numerical model for the evolution of the elemental abundances in the Galactic ISM, and found that the dust mass is linearly proportional to the ISM metallicity.
This might be expected since dust is primarily composed of metals.
Edmunds (2001) built a semi-analytical model for the dust evolution in galaxies, and found that the interstellar dust mass should be an approximately constant fraction of the heavy (metal) element mass in the ISM, consistent with Dwek (1998).
Very recently, Zafar \& Watson (2013) measured the metal-to-dust ratios of high redshift galaxies using GRBs and QSO absorbers.
Their results indicate that the metal-to-dust ratio is independent of galaxy type or age, redshift, or metallicity, and is very close to the value in the Local Group. 

Figure~\ref{fig:dtg_metal} shows the correlation between metallicity and dust-to-gas ratio for a number of nearby galaxies (Issa et al.\ 1990) and our three measurements.
We also show the data for the eight GRBs and the four quasar DLAs with metallicity measurements (see Table~\ref{tab:zafar13}).   
It appears that the dust-to-gas ratio of the lenses is linearly correlated with their metallicities, consistent with those nearby galaxies (Viallefond et al.\ 1982; Issa et al.\ 1990) and the theoretical models of Dwek (1998) and Edmunds (2001).
The slope of the linear correlation
\be\label{slope}
\frac{Z}{Z_\odot} = a \frac{E(B-V)}{N_H}
\ee (where the dust-to-gas ratio is in Galactic Units)
 is $a=1.21^{+0.57}_{-0.56}$ for the three lenses and $a=0.70^{+0.11}_{-0.12}$ after including the GRB and DLA data.  
The relative agreement between all the samples in Figure~\ref{fig:dtg_metal} suggests that there is little evolution in the metal-to-dust ratio with redshift, although there might be evolution in the metallicity $Z$ with redshift. 
As we see in Figure~\ref{fig:Metal_Z}, the GRBs at higher redshifts ($2.0<z<4.7$) seem have a lower average metallicity  than the three lenses at $z<1$.

Inoue (2003) constructed a semi-analytic model of the evolution of metals and dust in the ISM in order to study the evolution of the dust-to-metal ratio in galaxies.
His model predicts various evolutionary tracks in the metallicity---dust-to-gas ratio plane depending on the star-formation history (see Figure 3 of Inoue 2003), and this model fits the correlations observed for local galaxies well.  
We compare our three measurements to the Inoue (2003) predictions in Figure~\ref{fig:Inoue}.
The metallicity and dust-to-gas ratio of Q~2237$+$0305  agree with the evolutionary track corresponding to its redshifts $z=0.0395$ very well.
However, the dust-to-gas ratio of B~1152+199 is about a factor of 2 higher than the value predicted by the evolution track corresponding to its redshift $z=0.439,$ and is inconsistent with the model of Inoue (2003) by more than $3\sigma$.
The uncertainties in the metallicity and dust-to-gas ratio for SDSS~1004$+$4112 (redshift $0.68$) are too large for a meaningful comparison.
The dust-to-gas ratios of the high redshfit GRBs and DLAs are always significantly greater than the model predictions for their redshifts.
This implies that dust in the ISM in the high redshift galaxies forms very rapidly (faster than predicted by Inoue 2003), consistent with Zafar \& Watson (2013).
More lenses with direct metallicity measurements and smaller uncertainties are needed in order to test these ISM evolution models. 
\subsection{Dust-to-Gas Ratio of Gravitational Lenses, GRB Host Galaxies, and Quasar Foreground Absorbers}

Figure~\ref{fig:EBV_NH} shows the correlation between the differential color excess $\Delta E(B-V)$ and the differential absorption $\Delta N_H$ including both our new results and those from Dai \& Kochanek (2009).  
For comparison we also show a local stellar sample from Bohlin et al.\ (1978).
We show the lens dust-to-gas ratio as a function of redshift in Figure~\ref{fig:dtg_z}. 
We find an average dust-to-gas ratio for the lenses of
\be
\Bigg\langle\frac{E(B-V)}{N_H}\Bigg\rangle = \dtglens,
\ee with an intrinsic scatter of $\rm 0.3\, dex$ that is included in the model fits (see Table~\ref{tab:mdtg}).
The intrinsic scatter 
is larger than the Galactic value (Bohlin et al.\ 1978),  $\sim$$0.5 \times 10^{-22}\,\rm mag\, cm^{2}\, atoms^{-1}$ (about 30\%) or the value from Dai \& Kochnek (2009),  $\sim$$0.6 \times 10^{-22}\,\rm mag\, cm^{2}\, atoms^{-1}$ (or $\sim$$40\%$).
The average dust-to-gas ratio is slightly lower than the Galactic mean of $1.7 \times 10^{-22}\,\rm mag\, cm^{2}\, atoms^{-1}$ (Bohlin et al.\ 1978).
Dai \& Kochanek (2009) reported a mean dust-to-gas ratio,  $(1.5\pm0.5)\times 10^{-22}\,\rm mag\, cm^{2}\, atoms^{-1},$ consistent with the Galactic mean value, using  a smaller sample of 6 lenses with relatively large differential absorption.
In some ways, the consistency between the Dai \& Kochanek (2009) value and the Galactic value is puzzling, because several models predict lower dust-to-gas ratios for galaxies at high redshifts (e.g., Dav$\rm\acute{e}$ \& Oppenheimer 2007).
The lower dust-to-gas ratio of lens galaxies ($0.04<z<1.0$) reported in this paper is more consistent with predictions from evolution models (e.g., Dwek 1998; Edmunds 2001; Inoue 2003) and also agrees well with the mean dust-to-gas ratio of Mg II absorbers ($0.5<z<1.4$) of $(0.97 \pm 0.19)  \times 10^{-22}\,\rm mag\, cm^{2}\, atoms^{-1},$ found by M$\rm\acute{e}$nard \& Chelouche (2009)  (see also Rao et al.\ 2006 and M$\rm\acute{e}$nard et al.\ 2008).
  
We computed the mean dust-to-gas ratio of GRB afterglows and quasar foreground absorbers, and compared the various samples in Figures~\ref{fig:EBV_NH} and \ref{fig:dtg_z}, and in Table~\ref{tab:mdtg}.
The redshifts of the GRB and DLA sample ($z>1.9$ except for one DLA at redshift 0.72) are higher than the gravitational lens sample ($z<1$). 
The mean dust-to-gas ratio of GRB+DLA sample is $\dtgGRB$, about one third the value found for the lens sample. 
This suggests evolution of the dust-to-gas ratio with redshift, but this should be interpreted cautiously because the environments of these measurements are drastically different.
If we simply combine the lens, GRB, and DLA sample, we find a mean dust-to-gas ratio $\mdtg$, significantly lower than the Galactic value (at the $6.8\,\sigma$ significance). 
The intrinsic scatter is about $\rm 0.62\, dex$, larger than the scatter in the lens sample.
If the dust-to-metal ratios of high redshift ($z>0$) galaxies are very close to that of the Local group,
as indicated by the lens and/or the GRB+DLA sample (Figure~\ref{fig:dtg_metal}), then the lower mean dust-to-gas ratio of the high redshift ($z>0$) galaxies presumably implies the decline of metallicity with redshift. 

\section{CONCLUSION}

\begin{figure}
	\epsscale{0.8}
	\plotone{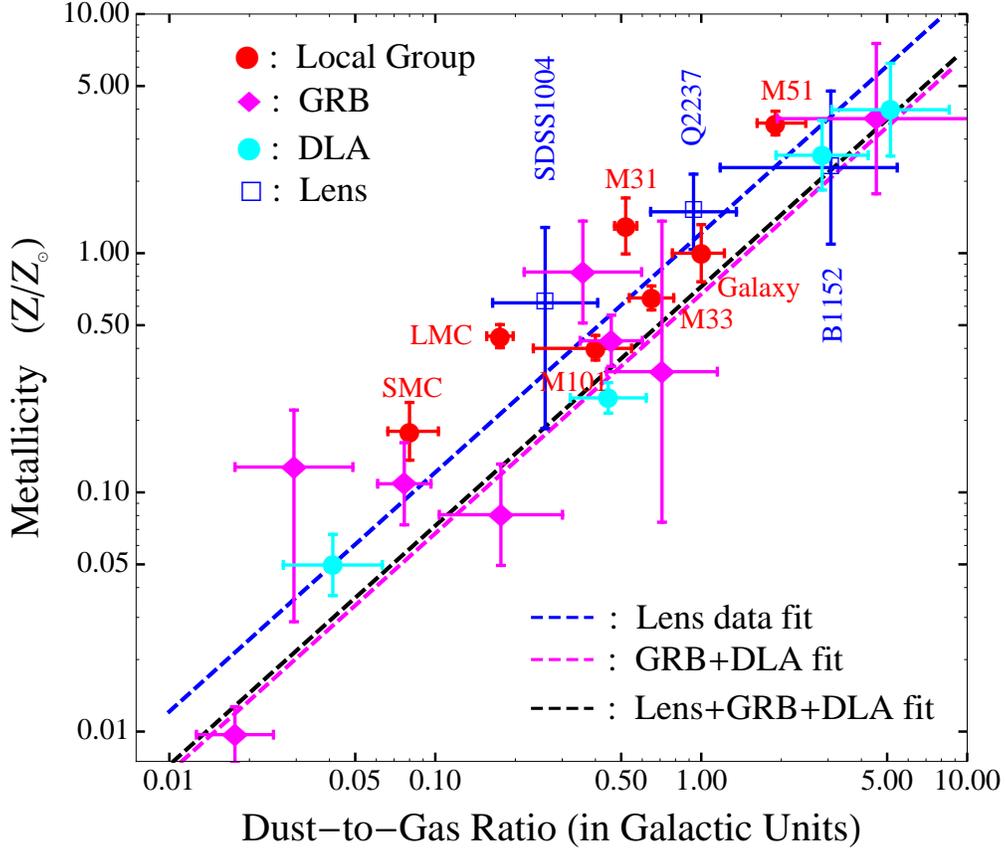}
	\caption{	Dust-to-gas ratio versus metallicity of the lenses (blue squares), GRBs (magenta diamonds), and DLAs (filled cyan circles), as compared to nearby galaxies. 
			The values for the Local Group  
			 are shown as filled red circles (Issa et al.\ 1990). 
	                The blue dashed line shows the best linear fit for the lenses with a slope $a=1.21^{+0.57}_{-0.56}$ (see Equation~(\ref{slope})).
	                The best linear fit to the GRB+DLA data and the  lens+GRB+DLA data  has a slope $a=0.67^{+0.12}_{-0.11}$ (the magenta dashed line) and $a=0.70^{+0.11}_{-0.12}$ (the black dashed line), respectively. 
	               All fits suggest a  linear correlation between dust-to-gas ratio and metallicity, or equivalently a constant metal-to-dust ratio. 
		       	   \label{fig:dtg_metal}} 
\end{figure}

\begin{figure}
	\epsscale{0.8}
	\plotone{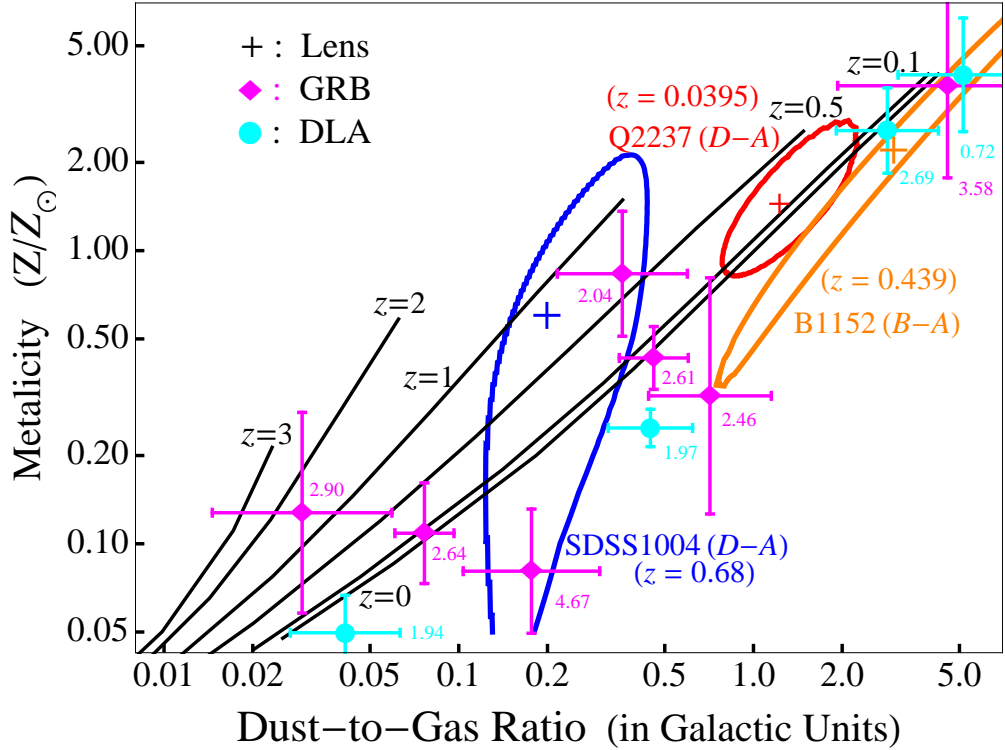}
	\caption{	Redshift evolution of the dust-to-gas ratio and metallicity of the lenses, GRBs and quasar Ly$\alpha$ absorbers.   
			The contours for the lenses are at the $68\%$ confidence limit for two parameters.
			The numbers in magenta and cyan are the redshifts of GRBs and DLAs, respectively.	
		      	The curves show the predicted redshift evolution of the metallicity---dust-to-gas ratio relation from Inoue (2003). 
		      	The metallicity and dust-to-gas ratio of Q~2237$+$0305 agree with the evolutionary track corresponding to its redshift $z=0.0395$ very well.
			The dust-to-gas ratio of B~1152+199 is a factor of $\sim$2 higher than the value predicted by the evolution track corresponding its redshift ($z=0.439$).
			The uncertainties for SDSS~1004$+$4112 (at redshift 0.68) are too large for a meaningful comparison.
		        The dust-to-gas ratios of GRBs and DLAs are significantly higher than the values predicted by the evolutional tracks corresponding to their redshifts, implying very rapid dust formation in the ISM of high redshift galaxies.
 		       	   \label{fig:Inoue}} 
\end{figure}

\begin{figure}
	\epsscale{0.8}
	\plotone{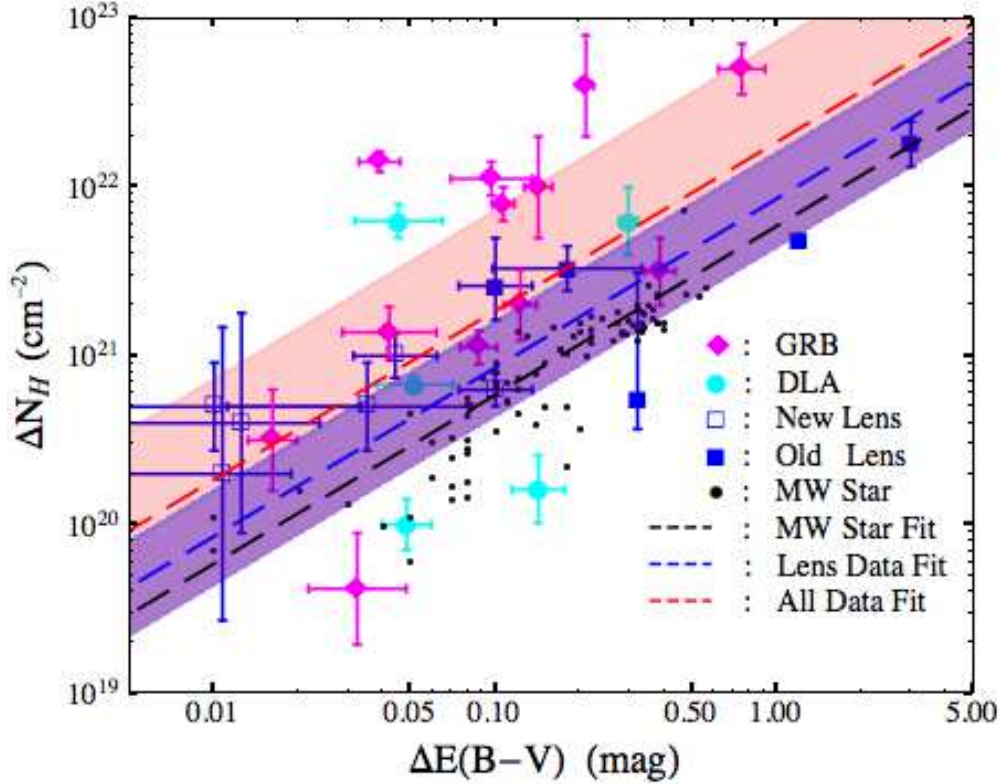}
	\caption{ 	Differential gas column density $\Delta N_H$ versus differential extinction $\Delta E(B-V)$ for gravitational lenses.
			Filled and open squares are lenses from Dai \& Kochanek (2009) and the new \chandra\ observations, respectively.	
			The smaller points are Galactic stars from Bohlin et al.\ (1978).	
			The magenta diamonds and filled cyan circles are respectively GRBs and DLAs from Zafar \& Watson (2013).  
			The blue and red dashed lines show the average dust-to-gas ratio of the lens data and the lens+GRB+DLA data, respectively.
			The widths of the intrinsic scattering of the lens data and the lens+GRB+DLA data are shown as blue and red shadowed areas, respectively. 
			The average Galactic dust-to-gas ratio is shown as black dashed line.
			The best fit dust-to-gas ratio ($1.17^{+0.41}_{-0.31}  \times 10^{-22}\rm mag\,cm^2\,atom^{-1}$) of lens galaxies is lower than the average Galactic value ($1.7\times 10^{-22}\rm mag\,cm^2\,atom^{-1}$),  suggesting that galaxies at higher redshift on average have lower metallicity.
			Including GRBs and DLAs will result in an even smaller mean dust-to-gas ratio ($0.54^{+0.19}_{-0.14}  \times 10^{-22}\rm mag\,cm^2\,atom^{-1}$) than the Galactic value (at the $6.8\,\sigma$ significance).
					       	   \label{fig:EBV_NH}} 
\end{figure}

\begin{figure}
	\epsscale{0.6}
	\plotone{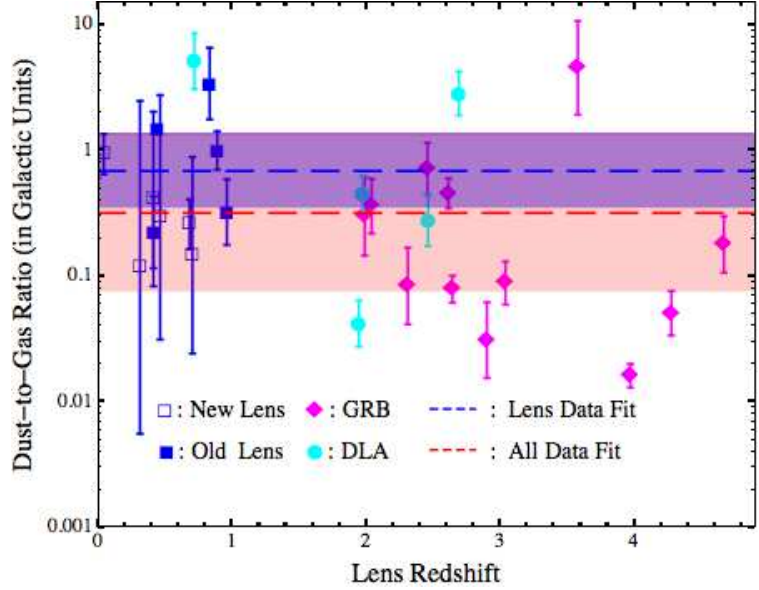}
	\caption{Dust-to-gas ratio in Galactic units ($1.7\times 10^{-22} \hbox{mag}\,\hbox{cm}^2\,\hbox{atom}^{-1}$) as a function of redshift $z.$
		      Filled and open blue squares are lenses from Dai \& Kochanek (2009) and the new results, respectively. 
		      	The blue and red dashed lines show the average dust-to-gas ratio of the lens data and the lens+GRB+DLA data, respectively.
			The widths of the intrinsic scattering of the lens data and the lens+GRB+DLA data are shown as blue and red shadowed areas, respectively. 
		        No significant evolution in the dust-to-gas ratio with redshift was detected. 
		       	   \label{fig:dtg_z}} 
\end{figure}

We summarize the main results of this paper as following: we homogeneously analyzed the \chandra\ X-ray observations of 10 gravitational lenses,  HE~0047$-$1756, QJ~0158$-$4325, SDSS~0246$-$0805, HE~0435$-$1223, SDSS~0924$+$0219, SDSS~1004+4112, HE~1104$-$1805,  PG~1115+080, Q~1355$-$2257, and Q~2237+0305, to measure the differential X-ray absorption between images, the metallicity, and the dust-to-gas ratio of the lens galaxies.
We detected differential absorption in all lenses except SDSS~0924$+$0219 and HE~1104$-1805$.
We successfully measured the gas phase metallicity of three lenses, Q~2237+0305, SDSS~1004+4112, and B~1152$+$199.
We estimated the dust-to-gas ratios of 11 lenses, 12 GRBs, and 5 QSO DLAs.
We find evidence for a linear correlation between the metallicity of the ISM in high redshift galaxies and its dust-to-gas ratio (Figure~\ref{fig:dtg_metal}).
This correlation is very similar to that of the Local Group, implying that the metal-to-dust ratios of high redshift ($z>0$) galaxies are very close to those of the Local Group.
Consequently, our results suggest very rapid dust formation in high redshift ($z>2$) galaxies (Figure~\ref{fig:Inoue}), consistent with what was recently found in Zafar \& Watson (2013).
The mean dust-to-gas ratio of the lens galaxies ($z<1$) is lower than the Galactic value, whereas that of the GRBs and DLAs ($z\gtrsim2$) is even lower (Figure~\ref{fig:EBV_NH}), which suggests cosmological metallicity evolution, \ie galaxies at higher redshift on average have lower metallicity.

We thank Akio Inoue for providing the data for plotting the model curves in Figure~\ref{fig:Inoue}.
We acknowledge support for this work provided by the National Aeronautics and Space Administration through Chandra Award Number GO0-11121ABC, GO1-12139ABC, GO2-13132ABC,  60014522 and AR2-13003X, issued by the Chandra X-ray Observatory Center, which is operated by the Smithsonian Astrophysical Observatory for and on behalf of the National Aeronautics Space Administration under contract NAS8-03060.
C.S.K. is supported by NSF grant AST-1009756.

\newpage

\begin{deluxetable}{lcccllc}
\tabletypesize{\scriptsize}
\tablecolumns{9}
\tablewidth{0pt}
\tablecaption{Basic Information \label{tab:lens-info}}
\tablehead{
\colhead{Lens}&
\colhead{Images}&
\colhead{$z_s$}&
\colhead{$z_l$\tablenotemark{a} }&
\colhead{ RA (J2000)}&
\colhead{ DEC (J2000) }&
\colhead{ Galactic $\rm N_H$\tablenotemark{b}  }\\
\colhead{}&
\colhead{}&
\colhead{source}&
\colhead{lens}&
\colhead{}&
\colhead{}&
\colhead{$(10^{22}\,{\rm cm}^{-2})$} 
 }
\startdata
HE~0047$-$1756    	  &  2  &  1.66      	&  0.41     	& 00:50:27.83 	& $-$17:40:08.8      & 0.0173 \\
QJ~0158$-$4325        & 2    & 1.29       	&  0.317   	& 01:58:41.44  	&  $-$43:25:04.2  & 0.0195     \\
SDSS~0246$-$0825  & 2    &  1.68      	&  0.724   	& 02:46:34.11 	&  $-$08:25:36.2    & 0.0332   \\ 
HE~0435$-$1223       &  4   &  1.689    	&  0.46   	& 04:38:14.9    	&  $-$12:17:14.4   & 0.0511   \\
SDSS~0924+0219     &  4    & 1.524    	&  0.39     & 09:24:55.87  	&    +02:19:24.9      & 0.0375     \\ 
SDSS~1004+4112     &  4   & 1.734    	&  0.68     	& 10:04:34.91  	&    +41:12:42.8      & 0.0111    \\ 
HE~1104$-$1805       &  2   &   2.32       	&  0.73     	& 11:06:33.45  	&  $-$18:21:24.2   & 0.0462  \\
PG~1115+080             &  4   &  1.72       	& 0.31      & 11:18:17.00  	&   +07:45:57.7      & 0.0356 \\
Q~1355$-$2257          & 2   &  1.37      	&  0.701 	& 13:55:43.38 	&  $-$22:57:22.9    & 0.0589   \\
Q~2237+0305             &   4 &  1.69       	&  0.0395 	&  22:40:30.34 	&    +03:21:28.8      & 0.0551   \\
\enddata
\tablenotetext{a}{ See the text for references.}
\tablenotetext{b}{The Galactic absorption is from Dickey \& Lockman (1990). }
\end{deluxetable}

\begin{deluxetable}{lrlr}
\tabletypesize{\scriptsize}
\tablecolumns{4}
\tablewidth{0pt}
\tablecaption{Log of the New \chandra\ Observations 
\label{tab:chandra_new}}
\tablehead{
\colhead{Lens}&
\colhead{$\rm Obs\_{ID}$}&
\colhead{Date }&
\colhead{Exp.\ (s)} 
 }
\startdata
Q~1355$-$2257          & 9514   &  2008 Mar 23      &  31,185   \\
\tableline
HE~0047$-$1756    	  &  9515  &  2008 Aug 24      &  33,667     \\
				  & 10576 & 2008 Sep 01       & 6,052    \\
\tableline
SDSS~0246$-$0825  & 9516  & 2007 Nov 23        &  49,518   \\
\tableline
			             &  363    & 2000 Jun 02        & 26,489      \\
				  & 1630   & 2000  Nov 03       & 9,826    \\	
			           & 7757   & 2008 Jan  31       & 28,761    \\
			          & 10730 & 2008 Nov 02       & 14,641    \\ 
PG~1115+080		 & 10795 & 2009 Feb 09       & 14,545    \\ 			          
			          & 10796 & 2009 Mar 27       & 14,557    \\ 
			          & 11857 & 2010 Feb 01       & 14,582    \\ 
			          & 12093 & 2010 Feb 07       & 14,583    \\ 
			          & 12094 & 2010 Feb 15       & 14,584    \\ 
\enddata
\end{deluxetable}


\begin{deluxetable}{llcllcl}
\tabletypesize{\tiny}
\tablecolumns{7}
\tablewidth{0pt}
\tablecaption{Spectral Fitting Results \label{tab:spectra}}
\tablehead{
\colhead{Quasar}&
\colhead{Model}&
\colhead{Image}&
\colhead{$\Gamma$}&
\colhead{$N_H$ }&
\colhead{Metallicity} &
\colhead{$\chi^2({\rm dof})$}\\
\colhead{}& 
\colhead{}&
\colhead{}&
\colhead{}&
\colhead{($10^{22}\,\rm cm^{-2}$)}&
\colhead{($Z_\odot$)}&
\colhead{}
 }
\startdata
Q~1355$-$2257           & $\rm wabs(zwabs(pow))$ & A  & $1.83\pm0.06$ & $0.08\pm0.05$   		& \nodata  &  1.09 (65)  \\
			             & $\rm wabs(zwabs(pow))$ & B  &   \nodata             & $0.04^{+0.08}_{-0.04}$ & \nodata  &  \nodata  \\
\tableline
HE~0047$-$1756    	   & $\rm wabs(zwabs(pow))$  & A  & $1.87\pm0.07$  & $0.06\pm0.04$                 & \nodata   &   1.20 (88)    \\
			             & $\rm wabs(zwabs(pow))$ & B  &   \nodata              & $0.00^{+0.02}_{-0.00}$  & \nodata   &   \nodata          \\
\tableline
HE~0047$-$1756    	   & $\rm wabs(zwabs(pow+zgauss))$  &  A  & $1.88\pm0.07$  & $0.05\pm0.03$                 & \nodata   &   1.16 (85)    \\
			             & $\rm wabs(zwabs(pow+zgauss))$  & B  &   \nodata              & $0.00^{+0.02}_{-0.00}$  & \nodata   &   \nodata          \\
\tableline
SDSS~0246$-$0825  & $\rm wabs(zwabs(pow))$  & A  & $2.13\pm0.08$  & $0.11\pm0.05$                 & \nodata  & 1.15 (61)  \\
			             & $\rm wabs(zwabs(pow))$ & B  &  \nodata              &  $0.00^{+0.01}_{-0.00}$  & \nodata  &  \nodata \\
\tableline
SDSS~0246$-$0825  & $\rm wabs(zwabs(pow+zgauss))$  & A  & $2.15\pm0.07$  & $0.10\pm0.05$                 & \nodata  & 1.00 (58)  \\
			             & $\rm wabs(zwabs(pow+zgauss))$ & B  &  \nodata              &  $0.00^{+0.01}_{-0.00}$  & \nodata  &  \nodata \\			             
\tableline
PG~1115+080              & $\rm wabs(zwabs(pow))$  &A1  	& $1.70\pm0.04$	& $0.17\pm0.02$  	& \nodata    &  1.31 (363)   \\
			             & $\rm wabs(zwabs(pow))$  &A2  	& \nodata       		& $0.25\pm0.03$  	& \nodata    &  \nodata  \\
			             & $\rm wabs(zwabs(pow))$  & B   	& \nodata       		& $0.09\pm0.02$ 	& \nodata    &  \nodata \\
		                       & $\rm wabs(zwabs(pow))$ & C  	& \nodata       		& $0.24\pm0.03$  	& \nodata    &  \nodata \\
\tableline
PG~1115+080              & $\rm wabs(zwabs(pow+zgauss))$  &A1  	& $1.62\pm0.05$	& $0.09\pm0.03$  	& \nodata    &  1.05 (354)   \\
			             & $\rm wabs(zwabs(pow+zgauss))$  &A2  	& \nodata       		& $0.14\pm0.03$  	& \nodata    &  \nodata  \\
			             & $\rm wabs(zwabs(pow+zgauss))$  & B   	& \nodata       		& $0.06\pm0.03$ 	& \nodata    &  \nodata \\
		                     & $\rm wabs(zwabs(pow+zgauss))$ & C  	& \nodata       		& $0.18\pm0.04$  	& \nodata    &  \nodata \\

\tableline
PG~1115+080\tablenotemark{a}              & $\rm wabs(zwabs^2(pow+zgauss))$  &A1  	& $1.62\pm0.05$	& $0.09\pm0.03$  	& \nodata    &  1.05 (353)   \\
			             & $\rm wabs(zwabs^2(pow+zgauss))$  &A2  	& \nodata       		& $0.14\pm0.03$  	& \nodata    &  \nodata  \\
			             & $\rm wabs(zwabs^2(pow+zgauss))$  & B   	& \nodata       		& $0.06\pm0.03$ 	& \nodata    &  \nodata \\
		                     & $\rm wabs(zwabs^2(pow+zgauss))$ & C  	& \nodata       		& $0.18\pm0.04$  	& \nodata    &  \nodata \\
\tableline
PG~1115+080\tablenotemark{b}      & $\rm wabs(zwabs^2(pow+zgauss))$  &A1  	& $1.58^{+0.07}_{-0.09}$	& $0.08^{+0.04}_{-0.08}$  	& \nodata    &  1.03 (237)   \\
			             			& $\rm wabs(zwabs^2(pow+zgauss))$  &A2  	& \nodata       			& $0.13^{+0.04}_{-0.09}$  	& \nodata    &  \nodata  \\
\tableline
PG~1115+080\tablenotemark{c}      & $\rm wabs(zwabs^2(pow+zgauss))$  &A1  	& $1.58^{+0.06}_{-0.08}$	& $<0.06$  				& \nodata    &  1.03 (236)   \\
			             			& $\rm wabs(zwabs^2(pow+zgauss))$  &A2  	& \nodata       			& $0.12^{+0.04}_{-0.05}$  	& \nodata    &  \nodata  \\

\tableline
QJ~0158$-$4325        &  $\rm wabs(zwabs(pow))$  &A  &  $1.93 \pm 0.06$    	&  $0.00^{+0.01}_{-0.00}$  & \nodata     &  0.93 (129)       \\ 
			             & $\rm wabs(zwabs(pow))$ & B  &  \nodata                    	&  $0.03^{+0.03}_{-0.02}$  & \nodata     & \nodata       \\
\tableline
HE~0435$-$1223       &   $\rm wabs(zwabs(pow))$ & A  & $1.92\pm0.10$       	&  $0.03\pm0.06$      	&  \nodata     &     1.08 (100)   \\
			             & $\rm wabs(zwabs(pow))$  & B  &  \nodata 		       	&  $0.01\pm0.04$      	&  \nodata     &    \nodata          \\
			             & $\rm wabs(zwabs(pow))$  & C  & $1.68\pm0.10$       	&  $0.01^{+0.07}_{-0.01}$&  \nodata     &   1.01 (84)         \\
		                       & $\rm wabs(zwabs(pow))$ & D  & \nodata                   	&  $0.00^{+0.09}_{-0.00}$&  \nodata     &   \nodata            \\
\tableline
SDSS~0924+0219     &  $\rm wabs(zwabs(pow))$   & A+C+D\tablenotemark{d}  & $2.23^{+0.07}_{-0.05}$ & $0.00^{+0.04}_{-0.00}$     & \nodata   &  0.91 (125)          \\ 
			             & $\rm wabs(zwabs(pow))$  & B             & $2.14^{+0.15}_{-0.14}$ & $0.00^{+0.07}_{-0.00}$     & \nodata   &  0.69 (27)    \\
\tableline
SDSS~1004+4112       & $\rm wabs(zwabs(pow))$  & A  & $1.99\pm0.03$      	& $0.04\pm0.03$   & \nodata     &  1.15 (659)         \\ 
			             & $\rm wabs(zwabs(pow))$  & B  & \nodata       		& $0.09\pm0.01$  &  \nodata     &  \nodata               \\
			             & $\rm wabs(zwabs(pow))$  & C  & \nodata       		& $0.09\pm0.01$  &  \nodata     &  \nodata                \\
		                      & $\rm wabs(zwabs(pow))$ & D  &  \nodata      		& $0.13\pm0.02$  &  \nodata      & \nodata                \\
\tableline
SDSS~1004+4112       &   $\rm wabs(zvphabs(pow))$ & A & $1.88\pm0.06$     &  $0.00^{+0.10}_{-0.00}$           & $0.62\pm0.60$        &  1.08 (277)     \\
		                       & $\rm wabs(zvphabs(pow))$ &  D & \nodata                   & $0.13\pm0.05$                          &  \nodata   &  \nodata             \\
\tableline
HE~1104$-$1805         &  $\rm wabs(zwabs(pow))$  & A  & $1.81\pm0.03$       &  $0.00^{+0.01}_{-0.00}$   &  \nodata   &  1.40 (48)  \\
			             &   $\rm wabs(zwabs(pow))$  & B  &  \nodata                   &  $0.00^{+0.01}_{-0.00} $  &  \nodata   &   \nodata   \\
\tableline
Q~2237+0305               & $\rm wabs(zwabs(pow))$ & A & $1.90\pm0.04$   & $0.07\pm0.01$    &  \nodata  & 1.30 (98)   \\
			             & $\rm wabs(zwabs(pow))$  &B & $1.93\pm0.08$                  & $0.08\pm0.02$     &  \nodata  & 1.17 (97)  \\
			             & $\rm wabs(zwabs(pow))$  &C & $2.04^{+0.10}_{-0.09}$   & $0.07\pm0.03$     &   \nodata  & 0.99 (75)  \\
		                       & $\rm wabs(zwabs(pow))$     & D& $1.94\pm0.07$              &  $0.14\pm0.02$    & \nodata  & 0.92 (128)  \\ 
\tableline
Q~2237+0305		     & $\rm wabs(zvphabs(pow))$ & A & $1.94\pm0.04$              &  $0.06\pm0.02$    & $1.49\pm0.54$   &  1.38 (236)   \\
		                       & $\rm wabs(zvphabs(pow))$ & D &  \nodata                          &  $0.11\pm0.03$    &  \nodata & \nodata  \\	
\tableline
B~1152+199                  & $\rm wabs(zwabs(pow))$  &A & $2.10\pm0.04$   & $0.002^{+0.010}_{-0.002}$    &  \nodata  & 1.04 (224)   \\
			              & $\rm wabs(zwabs(pow))$  &B &  \nodata               & $0.48\pm0.04$                          &  \nodata  & \nodata        \\
\tableline
B~1152+199		    & $\rm wabs(zvphabs(pow))$ & A & $2.10\pm0.05$        &  $0.00^{+0.01}_{-0.00}$    & $2.28\pm1.68$   &  1.04 (223)   \\
		                    & $\rm wabs(zvphabs(pow))$ & B &   \nodata                    &  $0.23^{+0.18}_{-0.18}$                  &  \nodata & \nodata  \\		 	                       
\enddata
\tablenotetext{a}{We fit the spectra of PG~1115+080 with power law plus  Gaussian emission lines modified by absorption at the source, the lens, and the Milky Way. 
 			   We assume the same source absorption column density $N_H^{\rm Src}$ for all lens images. 
			   We detect no source absorption and constrain the source absorption $N_H^{\rm Src}<0.08\times10^{22}\,\rm cm^{-2}.$  }
\tablenotetext{b}{We fit the spectra of images A1 and A2 jointly including the source absorption.
  			   We assume the same source absorption for the two images. 
			   We detect no source absorption and constrain $N_H^{\rm Src}<0.47\times10^{22}\,\rm cm^{-2}.$  }
\tablenotetext{c}{We fit the spectra of image A1 and A2 jointly, assuming independent source absorption for the images. 
			   We measure $N_{H,\,A1}^{\rm Src}=0.43^{+0.18}_{-0.35}\times10^{22}\,\rm cm^{-2}$ and  $N_{H,\,A2}^{\rm Src}<0.2\times10^{22}\,\rm cm^{-2}.$ }
\tablenotetext{d}{Images C and D of SDSS~0924+0219 are faint and poorly resolved, we extract a single spectrum containing the images A, C, and D.}
\end{deluxetable}


\begin{deluxetable}{lccccc}
\tabletypesize{\scriptsize}
\tablecolumns{6}
\tablewidth{0pt}
\tablecaption{The Iron Line Emission Detection  \label{tab:iron}}
\tablehead{
\colhead{Lens}&
\colhead{Image}&
\colhead{$E_{\rm line}$ (keV) }&
\colhead{$\sigma$ (keV)} &
\colhead{EW (keV)\tablenotemark{b} } &
\colhead{Significance} 
 }
\startdata
HE~0047$-$1756       	&  A					& $5.69^{+0.20}_{-0.18}$       &  $0.30^{+0.17}_{-0.30}$ &   $0.56^{+0.24}_{-0.24}$ & $98.5\%$      \\ 
			       		&  B\tablenotemark{a}       & \nodata       &  \nodata          &   \nodata &  \nodata      \\
\tableline
SDSS~0246$-$0825       	&  A					& $5.73^{+0.12}_{-0.12}$       &  $0.28^{+0.11}_{-0.10}$ &   $0.72^{+0.27}_{-0.24}$ & $99.8\%$      \\ 
			       		&  B\tablenotemark{a}       & \nodata       &  \nodata          &   \nodata &  \nodata      \\
\tableline 
 PG~1115+080       	&  A1    & $5.43^{+0.30}_{-0.39}$       &  $1.34^{+0.36}_{-0.34}$ &   $1.80^{+0.49}_{-0.49}$ & $99.96\%$      \\ 
			       	&  A2     & $5.11^{+0.25}_{-0.28}$      & $1.40^{+0.27}_{-0.23}$    & $2.56^{+0.70}_{-0.60}$  & $99.999\%$    \\
			       	&  B\tablenotemark{a}       & \nodata       &  \nodata          &   \nodata &  \nodata      \\ 
	     			&  C     &  $5.70^{+0.10}_{-0.10}$        & $0.23^{+0.19}_{-0.15}$   & $0.79^{+0.24}_{-0.24}$ &   $99.84\%$  \\ 
\enddata
\tablenotetext{a}{Where an image had no detection the entries are left empty.}
\tablenotetext{b}{EW is the rest frame equivalent width of the iron line component.}
\end{deluxetable}



\begin{deluxetable}{llllccc}
\tabletypesize{\scriptsize}
\tablecolumns{7}
\tablewidth{0pt}
\tablecaption{The Dust-To-Gas Ratios of High Redshift ($z>0$) Lens Galaxies. 
\label{tab:dtg}}
\tablehead{
\colhead{Lens} &
\colhead{Type} & 
\colhead{$z_l$} &
\colhead{Pair} &
\colhead{$\Delta N_H$} & 
\colhead{$\Delta E(B-V)$} &
\colhead{$E(B-V)/N_H$} \\
\colhead{} &
\colhead{} &
\colhead{} &
\colhead{}&
\colhead{($10^{22}\hbox{atoms }\hbox{cm}^{-2}$)} &
\colhead{(mag)} &
\colhead{$10^{-22}\hbox{mag cm}^2\hbox{atoms}^{-1}$}
}
\startdata
New Measurements   &    &     &   &     &    &            \\
\tableline
QJ~0158$-$4325             & Elliptical   	&  $0.317$                &  B, A     	& $0.028\pm0.023$ 		& $-0.002\pm0.03\>\>\>\>\>\> $      	& microlensing  \\
Q~1355$-$2257               & Elliptical 	&  $0.701$                &  A, B     	& $0.04\pm0.06$      		& $0.012\pm0.013$        	      	& $0.25\pm0.45$ \\
SDSS~0246$-$0825       & Elliptical   	&  $0.724$                &  A, B     	& $0.10\pm0.05$      		& $-0.012\pm0.017\>\>\>\>  	$     	&  microlensing \\
SDSS~1004$+$4112      & Cluster      	&  $0.68$      		&  D, A     	& $0.10\pm0.03$      		& $0.044\pm0.015   	$      	& $ 0.44\pm0.20   $ \\
HE~0047$-$1756             & Elliptical   	&  $0.41 $     		&  A, B      & $0.05  \pm0.03  $   	& $0.035\pm0.052    $  		& $0.70\pm1.12$ \\
HE~0435$-$1223             & S0             	&  $0.46 $     		&  A, B      & $0.02  \pm0.04  $   	& $0.011\pm0.010       	$  		& $0.50\pm1.12    $ \\
HE~1104$-$1805             & Elliptical  	&   $0.729$   		&  A, B     	& $<0.015$    & $-0.07\pm0.01\>\>\>\>	$        	&  microlensing \\
PG~1115+080                   & Elliptical  	&  $0.31 $     		&  A2, A1  & $0.05  \pm0.03  $   	& $0.01\pm0.03	$ 	& $0.20\pm0.61    $ \\
Q~2237$+$0305               & Spiral      	&  $0.0395$  		&  D, A     	& $0.063\pm0.014$ 		& $0.10\pm0.03	$   	& $1.59\pm0.59$ \\
B~1152+199 (vary $Z$)              	& Elliptical  	&  $0.439$   		& B, A  	&  $0.23\pm0.18$             & $1.20\pm0.05	$  	& $5.22\pm 4.10$ \\
\tableline
Dai \& Kochanek (2009) Samples   &    &     &   &     &    &            \\
\tableline
SBS~0909+523        & Elliptical  &  $0.83$     & B, A  &  $0.055^{+0.095}_{-0.022}$ & $0.32\pm0.01$ & $5.8^{+3.9}_{-3.7}$ \\
FBQS~0951+2635   & Elliptical  &  $0.26$     & B, A  &  $0.49^{+0.49}_{-0.41}$        & $-0.12\pm0.02\>\>\>\>$ &  microlensing \\
B~1152+199 ($Z\equiv Z_\odot$)             & Elliptical  &  $0.439$   & B, A  &  $0.48\pm0.04$                       & $1.20\pm0.05$  & $2.5\pm 0.2$ \\
MG~0414+0534       & Elliptical &   $0.9584$ & A, B  &  $0.33\pm0.10$                       & $0.18\pm0.11$  & $0.55\pm0.33$ \\
B~1600+434             & Spiral      &   $0.41$      & B, A  & $0.26^{+0.17}_{-0.12}$         & $0.10\pm0.03$  & $0.38\pm0.25$ \\
PKS~1830$-$211    & Spiral      &   $0.886$    & B, A  & $1.8^{+0.5}_{-0.6}$                & $3.00\pm0.13$  & $1.7\pm0.6$ \\
\enddata
\tablecomments{The extinction references are: Kochanek et al.\ (2007) for QJ~0158$-$4325, Q~1355$-$2257, SDSS~1004+4112, and HE~0435$-$1223; Inada et al.\ (2005) for SDSS~0246$-$0825; Wisotzki et al.\ (2004) for HE~0047$-$1756; Falco et al.\ (1999) for HE~1104$-$1805, PG~1115+080, and Q~2237+0305.}

\end{deluxetable}


\begin{deluxetable}{llcc}
\tabletypesize{\scriptsize}
\tablecolumns{4}
\tablewidth{0pt}
\tablecaption{Metallicity Measurements \label{tab:metallicity}}
\tablehead{
\colhead{Lens}&
\colhead{$z_l$}&
\colhead{Metallicity }&
\colhead{$E(B-V)/N_H$}\\
\colhead{}&
\colhead{}&
\colhead{$(Z/Z_\odot)$}&
\colhead{$(10^{-22}\hbox{mag cm}^2\hbox{ atoms}^{-1})$} 
 }
\startdata
Q~2237+0305               	&  0.0395 	&  $1.49\pm0.54$       &  $1.59\pm0.59$   \\
B1152+199        		&  0.439   	&  $2.28\pm1.68$       &  $5.22\pm4.10$      \\
SDSS~1004+4112      	&  0.68      	&  $0.62\pm0.60$       &  $0.44\pm0.20$         \\ 
\enddata
\end{deluxetable}


\begin{deluxetable}{lccccc}
\tabletypesize{\scriptsize}
\tablecolumns{6}
\tablewidth{0pt}
\tablecaption{ Dust-to-Gas Ratios of  GRBs and Quasar DLAs. 
			   \label{tab:zafar13}}
\tablehead{
\colhead{GRB}&
\colhead{redshift }&
\colhead{$\rm \log N_H$}&
\colhead{$E(B-V)$ }&
\colhead{$E(B-V)/N_H$}&
\colhead{$Z/Z_{\odot}$}\\
\colhead{}&
\colhead{}&
\colhead{}&
\colhead{(mag)}&
\colhead{$(10^{-22}\hbox{mag cm}^2\hbox{ atoms}^{-1})$}&
\colhead{}
 }
\startdata
000926             &  2.038	&  $21.30\pm0.21$ 	&  $0.12\pm0.02$   		&   $0.61\pm0.31$  &  $0.83\pm0.41$	\\
030226		&  1.987   &  $20.50\pm0.30$ 	&  $0.016\pm0.003$   	&   $0.51\pm0.37$  &  \nodata 	\\
050401       	&  2.899	&  $22.60\pm0.30$	&  $0.21\pm0.01$            	&   $0.05\pm0.04$  &  $0.13\pm0.12$	\\
050505      	&  4.275	&  $22.05\pm0.10$  	&  $0.10\pm0.03$           	&   $0.09\pm0.03$  &   \nodata	\\ 
050730   		&  3.969	&  $22.15\pm0.06$   &  $0.04\pm0.01$            &   $0.03\pm0.01$  &  $0.010\pm0.003$	\\ 
050820A		&  2.612   	&  $21.05\pm0.10$	&  $0.09\pm0.01$		&   $0.78\pm0.21$  &  $0.43\pm0.11$	\\
070506		&  2.308  	&  $22.00\pm0.30$	&  $0.14\pm0.02$		&   $0.14\pm0.10$  &  \nodata	\\
070802		&  2.455   &  $21.50\pm0.20$	&  $0.38\pm0.05$		&   $1.21\pm0.58$  &  $0.32^{+0.46}_{-0.32} $	\\
080210		&  2.641	&  $21.90\pm0.10$	&  $0.11\pm0.01$		&   $0.13\pm0.03$  &  $0.11\pm0.04$	\\
080607		&  3.037	&  $22.70\pm0.15$	&  $0.75\pm0.14$		&   $0.15\pm0.06$  &   \nodata	\\
090323		&  3.577	&  $19.62\pm0.33$	&  $0.03\pm0.01$		&   $7.74\pm6.64$  &  $3.65\pm2.64$	\\
100219A		&  4.667	&  $21.14\pm0.15$	&  $0.04\pm0.02$		&   $0.30\pm0.16$  &  $0.08\pm0.04$	\\
\tableline
QSO			& redshift &	$\rm \log N_H$ &  $E(B-V)$			& $E(B-V)/N_H$ &  $Z/Z_{\odot}$	\\
\tableline
0016$-$0012	& 1.973   	&  $20.83\pm0.05$	&  $0.05\pm0.02$		&   $0.76\pm0.25$  &  $0.25\pm0.04$	\\
1157+6135	&  2.459	&  $21.80\pm0.20$	&  $0.30\pm0.02$		&   $0.47\pm0.22$  &  \nodata	\\
1159+0112	&  1.944	&  $21.80\pm0.10$	&  $0.05\pm0.02$		&   $0.07\pm0.03$  &  $0.050\pm0.015$	\\
1237+0647	&  2.690	&  $20.00\pm0.15$	&  $0.05\pm0.01$		&   $4.84\pm1.93$  &  $2.57\pm0.86$	\\
1323$-$0021	&  0.716	&  $20.21\pm0.20$	&  $0.14\pm0.03$		&   $8.75\pm4.45$  &  $3.98\pm1.78$	\\
\enddata
\tablecomments{We compiled a subsample of GRBs and DLAs from Table~1 of Zafar \& Watson (2013) by choosing only those systems with both absorption and extinction measurements (including uncertainty estimates).
			We  converted  $A_V$  to color excess $E(B-V)$ assuming Galactic extinction law with $R_V=3.1.$
			 The metallicity was estimated using the column density of the metal element Sulfur or Zinc, i.e., $N_{S}$ or $N_{Zn}$.  
			The references for GRBs are: 
				Chen et al.\ (2007) and Starling et al.\ (2007) for 000926;
				Shin et al.\ (2006) and Schady et al.\ (2011) for 030226;
				Watson et al.\ (2006) and Zafar et al.\ (2011) for 050401;
		  		Berger et al.\ (2006) and Hurkett et al.\ (2006) for 050505;	
				Chen et al.\ (2005b) and Zafar et al.\ (2011) for 050730;	
				Ledoux et al.\ (2009) and Schady et al.\ (2012) for 050820A;		
				Fynbo et al.\ (2009) and Zafar et al.\ (2011) for 070506;
				El$\acute{\i}$asd$\rm\acute{o}$ttir et al.\ (2009) and Zafar et al.\ (2011) for 070802;
				De Cia et al.\ (2011) and Zafar et al.\ (2011) for 080210;
				Prochaska et al.\ (2009) and Zafar et al.\ (2011) for 080607;
				Savaglio et al.\ (2012) and Schady et al.\ (2011) for 090323;
				and Th$\rm\ddot{o}$ne et al.\ (2013) for100219.	
			    The references for QSO DLAs are:  Vladilo et al.\ (2006) for 0016$-$0012, 1159+0112, and 1323$-$0021; Wang et al.\ (2012) for 1157+6135; Noterdaeme et al.\ (2010) for 1237+0647.   }
\end{deluxetable}


\begin{deluxetable}{cccccc}
\tabletypesize{\scriptsize}
\tablecolumns{6}
\tablewidth{0pt}
\tablecaption{Mean Dust-to-Gas Ratios of High Redshift Galaxies \label{tab:mdtg}}
\tablehead{
\colhead{Data}&
\colhead{sample size}&
\colhead{redshift}&
\colhead{$E(B-V)/N_H$}&
\colhead{$\log \sigma$\tablenotemark{a}}&
\colhead{$(\sigma_{\rm low},\sigma_{\rm up})$\tablenotemark{b} }\\
\colhead{}&
\colhead{}&
\colhead{}&
\colhead{$(10^{-22}\hbox{mag cm}^2\hbox{ atoms}^{-1})$}&
\colhead{(dex)}&
\colhead{$(10^{-22}\hbox{mag cm}^2\hbox{ atoms}^{-1})$}
 }
\startdata
Lens (old)                  	&  6	& $0.04<z<0.96$	&  $1.63^{+0.63}_{-0.46}$ 	& $0.27$	&  $(0.88,3.03)$            \\
Lens (new) 			& 11 & $0.04<z<0.96$	&  $1.17^{+0.41}_{-0.31}$ 	& $0.30$	&  $(0.60,2.31)$            \\
GRB+DLA        			& 17	& $0.72<z<4.67$	&  $0.39^{+0.19}_{-0.14}$ 	& $0.68$	&  $(0.09,1.90)$             \\
Lens (old)+GRB+DLA      	& 23	& $0.04<z<4.67$	&  $0.56^{+0.22}_{-0.17}$ 	& $0.66$  &  $(0.12,2.60)$               \\ 
Lens (new)+GRB+DLA   	& 28	& $0.04<z<4.67$	&  $0.54^{+0.19}_{-0.14}$ 	& $0.62$  &  $(0.14,2.26)$                 \\ 
\enddata
\tablecomments{The old sample of lenses is from Dai \& Kochanek (2009). 
			The (enlarged) new sample of lenses is from the current paper.
			 The GRB and QSO DLAs data are selected from Table 1 of Zafar \& Watson (2013), see Table~\ref{tab:zafar13}.}
\tablenotetext{a}{The intrinsic scatter in the mean dust-to-gas ratio.}
\tablenotetext{b}{The range of the intrinsic scatter. }
\end{deluxetable}

\end{document}